\documentclass[superscriptaddress,prd,showpacs,preprintnumbers,amsmath,twocolumn,amssymb,bibnotes,altaffilletter,floatfix,epsfig]{revtex4}

\def\thedirectory{.}

\def\cosz{$\cos{\theta}_{\rm zenith}$}
\def\theaddress{\@arabic\c@address}
\def\anum@address{Address \theaddress.}

\def\thetitle{Measurement of the Cosmic Ray and Neutrino-Induced Muon Flux at the Sudbury Neutrino Observatory }
\def\livetime{1229.30}
\def\livetimeerr{0.03}
\def\livetimesection{The data included in this analysis were collected during all three SNO operation phases. During the initial phase, data were collected from November 2, 1999 until May 28, 2001. During the second phase of SNO, data were recorded between July 26, 2001 and August 28, 2003.  The third and final phase of SNO operations collected data between November 27, 2004 and November 28, 2006.~}
\def\downwardefficiency{$99.2\%$~}
\def\upwardefficiency{$98.0\%$}
\def\luminosity{$2.30\times 10^{14}$ cm$^{2}$~s}

\def\rockrate{$124.4 \pm 6.5$~}
\def\waterrate{$9.0 \pm 0.5$~}
\def\internalrate{$3.1 \pm 0.8$~}
\def\nuerate{$1.9 \pm 0.3$~}
\def\signalrate{$138.4 \pm 7.3$~}
\def\cosmicrate{$1.1 \pm 1.2$~}
\def\instrumentalrate{$ 0.3 \pm 0.2$~}
\def\backgroundrate{$1.4 \pm 1.2$~}
\def\totalrate{$139.8 \pm 7.4$~}
\def\hrockrate{$43.2 \pm 2.3$~}
\def\hwaterrate{$2.8 \pm 0.2$~}
\def\hinternalrate{$1.0 \pm 0.3$~}
\def\hnuerate{$0.7 \pm 0.1$~}
\def\hsignalrate{$47.8 \pm 2.5$~}
\def\hcosmicrate{$1.1 \pm 1.2$~}
\def\hinstrumentalrate{$ 0.1 \pm 0.2$~}
\def\hbackgroundrate{$1.2 \pm 1.2$~}
\def\htotalrate{$49.0 \pm 2.8$~}
\def\downwardevents{76749~}
\def\upwardevents{514~}
\def\dataupward{152.7~}

\def\hupwardevents{201~}
\def\datahupward{59.7~}
\def\totalhflux{$3.31 \pm 0.23 {\rm (stat.)} \pm 0.13 {\rm (sys.)} \times 10^{-13}$}
\def\dailyrate{$62.9 \pm 0.2$~}
\def\totalvflux{$2.10 \pm 0.12 {\rm (stat.)} \pm 0.08 {\rm (sys.)} \times 10^{-13}$}
\def\flux{$(3.31 \pm 0.01 {\rm~(stat.)} \pm 0.09 {\rm ~(sys.)}) \times 10^{-10}~\mu$/s/cm$^{2}$} 
\def\normalization{$1.22 \pm 0.10$}
\def\normalizationconst{$1.22 \pm 0.09$}
\def\normalizationnoosc{$1.09\pm0.08$}
\def\msqatm{$2.6\times10^{-3}$~eV$^2$}
\def\thetaatm{$1.00$}
\def\CL{99.8\%~}

\usepackage{color}      
\usepackage{graphicx}
\usepackage{subfigure}

\begin{document}

\title{\thetitle}


%
\newcommand{\alta}{Department of Physics, University of 
Alberta, Edmonton, Alberta, T6G 2R3, Canada}
\newcommand{\ubc}{Department of Physics and Astronomy, University of 
British Columbia, Vancouver, BC V6T 1Z1, Canada}
\newcommand{\bnl}{Chemistry Department, Brookhaven National 
Laboratory,  Upton, NY 11973-5000}
\newcommand{\carleton}{Ottawa-Carleton Institute for Physics, Department of Physics, Carleton University, Ottawa, Ontario K1S 5B6, Canada}
\newcommand{\uog}{Physics Department, University of Guelph,  
Guelph, Ontario N1G 2W1, Canada}
\newcommand{\lu}{Department of Physics and Astronomy, Laurentian 
University, Sudbury, Ontario P3E 2C6, Canada}
\newcommand{\lbnl}{Institute for Nuclear and Particle Astrophysics and 
Nuclear Science Division, Lawrence Berkeley National Laboratory, Berkeley, CA 94720}
\newcommand{\lbla}{ Lawrence Berkeley National Laboratory, Berkeley, CA}
\newcommand{\lanl}{Los Alamos National Laboratory, Los Alamos, NM 87545}
\newcommand{\llnl}{Lawrence Livermore National Laboratory, Livermore, CA}
\newcommand{\lanla}{Los Alamos National Laboratory, Los Alamos, NM 87545}
\newcommand{\oxford}{Department of Physics, University of Oxford, 
Denys Wilkinson Building, Keble Road, Oxford OX1 3RH, UK}
\newcommand{\penn}{Department of Physics and Astronomy, University of 
Pennsylvania, Philadelphia, PA 19104-6396}
\newcommand{\queens}{Department of Physics, Queen's University, 
Kingston, Ontario K7L 3N6, Canada}
\newcommand{\uw}{Center for Experimental Nuclear Physics and Astrophysics, 
and Department of Physics, University of Washington, Seattle, WA 98195}
\newcommand{\uta}{Department of Physics, University of Texas at Austin, Austin, TX 78712-0264}
\newcommand{\triumf}{TRIUMF, 4004 Wesbrook Mall, Vancouver, BC V6T 2A3, Canada}
\newcommand{\ralimp}{Rutherford Appleton Laboratory, Chilton, Didcot OX11 0QX, UK}
\newcommand{\iusb}{Department of Physics and Astronomy, Indiana University, South Bend, IN}
\newcommand{\fnal}{Fermilab, Batavia, IL}
\newcommand{\uo}{Department of Physics and Astronomy, University of Oregon, Eugene, OR}
\newcommand{\hu}{School of Engineering, Hiroshima University, Hiroshima, Japan}
\newcommand{\slac}{Stanford Linear Accelerator Center, Menlo Park, CA}
\newcommand{\mac}{Department of Physics, McMaster University, Hamilton, ON}
\newcommand{\doe}{US Department of Energy, Germantown, MD}
\newcommand{\lund}{Department of Physics, Lund University, Lund, Sweden}
\newcommand{\mpi}{Max-Planck-Institut for Nuclear Physics, Heidelberg, Germany}
\newcommand{\uom}{Ren\'{e} J.A. L\'{e}vesque Laboratory, Universit\'{e} de Montr\'{e}al, Montreal, PQ}
\newcommand{\cwru}{Department of Physics, Case Western Reserve University, Cleveland, OH}
\newcommand{\pnnl}{Pacific Northwest National Laboratory, Richland, WA}
\newcommand{\uc}{Department of Physics, University of Chicago, Chicago, IL}
\newcommand{\mitt}{Laboratory for Nuclear Science, Massachusetts Institute of Technology, Cambridge, MA 02139}
\newcommand{\ucsd}{Department of Physics, University of California at San Diego, La Jolla, CA }
\newcommand{	\lsu	}{Department of Physics and Astronomy, Louisiana State University, Baton Rouge, LA 70803}
\newcommand{\imp}{Imperial College, London SW7 2AZ, UK}
\newcommand{\uci}{Department of Physics, University of California, Irvine, CA 92717}
\newcommand{\ucia}{Department of Physics, University of California, Irvine, CA}
\newcommand{\suss}{Department of Physics and Astronomy, University of Sussex, Brighton  BN1 9QH, UK}
\newcommand{	\lifep	}{Laborat\'{o}rio de Instrumenta\c{c}\~{a}o e F\'{\i}sica Experimental de
Part\'{\i}culas, Av. Elias Garcia 14, 1$^{\circ}$, 1000-149 Lisboa, Portugal}
\newcommand{\hku}{Department of Physics, The University of Hong Kong, Hong Kong.}
\newcommand{\aecl}{Atomic Energy of Canada, Limited, Chalk River Laboratories, Chalk River, ON K0J 1J0, Canada}
\newcommand{\nrc}{National Research Council of Canada, Ottawa, ON K1A 0R6, Canada}
\newcommand{\princeton}{Department of Physics, Princeton University, Princeton, NJ 08544}
\newcommand{\birkbeck}{Birkbeck College, University of London, Malet Road, London WC1E 7HX, UK}
\newcommand{\snoi}{SNOLAB, Sudbury, ON P3Y 1M3, Canada}
\newcommand{\uba}{University of Buenas Aires, Argentina}
\newcommand{\hvd}{Department of Physics, Harvard University, Cambridge, MA}
\newcommand{\pny}{Goldman Sachs, 85 Broad Street, New York, NY}
\newcommand{\pnv}{Remote Sensing Lab, PO Box 98521, Las Vegas, NV 89193}
\newcommand{\psis}{Paul Schiffer Institute, Villigen, Switzerland}
\newcommand{\liverpool}{Department of Physics, University of Liverpool, Liverpool, UK}
\newcommand{\uto}{Department of Physics, University of Toronto, Toronto, ON, Canada}
\newcommand{\uwisc}{Department of Physics, University of Wisconsin, Madison, WI}
\newcommand{\psu}{Department of Physics, Pennsylvania State University,
     University Park, PA}
\newcommand{\anl}{Deparment of Mathematics and Computer Science, Argonne
     National Laboratory, Lemont, IL}
\newcommand{\cornell}{Department of Physics, Cornell University, Ithaca, NY}
\newcommand{\tufts}{Department of Physics and Astronomy, Tufts University, Medford, MA}
\newcommand{\ucd}{Department of Chemical Engineering and Materials Science, University of California, Davis, CA}
\newcommand{\unc}{Department of Physics, University of North Carolina, Chapel Hill, NC}
\newcommand{\dresden}{Institut f\"{u}r Kern- und Teilchenphysik, Technische Universit\"{a}t Dresden,  01069 Dresden, Germany}
\newcommand{\fargo}{Business Direct, Wells Fargo, San Francisco, CA}
\newcommand{\ucol}{Physics Department, University of Colorado at Boulder, Boulder, CO}
\newcommand{\utah}{University of Utah Department of Physics, Salt Lake City, Utah}
\newcommand{\ucsb}{Department of Physics, University of California Santa Barbara, Santa Barbara, CA }
\newcommand{\cern}{CERN (European Laboratory for Particle Physics), Geneva, Switzerland}


\affiliation{\alta}
\affiliation{\ubc}
\affiliation{\bnl}
\affiliation{\carleton}
\affiliation{\uog}
\affiliation{\lu}
\affiliation{\lbnl}
\affiliation{\lifep}
\affiliation{\lanl}
\affiliation{\lsu}
\affiliation{\mitt}
\affiliation{\oxford}
\affiliation{\penn}
\affiliation{\queens}
\affiliation{\ralimp}
\affiliation{\snoi}
\affiliation{\uta}
\affiliation{\triumf}
\affiliation{\uw}

\author{B.~Aharmim}\affiliation{\lu}
\author{S.N.~Ahmed}\affiliation{\queens}
\author{T.C.~Andersen}\affiliation{\uog}
\author{A.E.~Anthony}\affiliation{\uta}
\author{N.~Barros}\affiliation{\lifep}
\author{E.W.~Beier}\affiliation{\penn}
\author{A.~Bellerive}\affiliation{\carleton}
\author{B.~Beltran}\affiliation{\alta}\affiliation{\queens}
\author{M.~Bergevin}\affiliation{\lbnl}\affiliation{\uog}
\author{S.D.~Biller}\affiliation{\oxford}
% ...
			
\collaboration{SNO Collaboration}
\noaffiliation

\begin{abstract}
Results are reported on the measurement of the atmospheric neutrino-induced muon flux at a depth of 2 kilometers below the Earth's surface from 1229 days of operation of the Sudbury Neutrino Observatory (SNO).  By measuring the flux of through-going muons as a function of zenith angle, the SNO experiment can distinguish between the oscillated and un-oscillated portion of the neutrino flux.  A total of \upwardevents muon-like events are measured between $-1 \le $\cosz$ \le 0.4$ in a total exposure of \luminosity. The measured flux normalization is \normalizationconst~times the Bartol three-dimensional flux prediction.  This is the first measurement of the neutrino-induced flux where neutrino oscillations are minimized.  The zenith distribution is consistent with previously measured atmospheric neutrino oscillation parameters.  The cosmic ray muon flux at SNO with zenith angle \cosz $> 0.4$ is measured to be \flux.\\
\end{abstract}


\date{\today}

\maketitle

\section{Introduction}
\label{sec:Intro}

Atmospheric neutrinos are produced from the decay of charged mesons created by the interactions of primary cosmic rays with the Earth's atmosphere.  Atmospheric neutrinos can be detected either via direct interactions within the fiducial volume of a given detector or indirectly from the observation of high-energy muons created via the charged current interaction $\nu_\mu+ N \rightarrow \mu + X$ on materials that surround the detector.  Although the latter process produces muons propagating at all zenith angles, overhead portions of the sky are typically dominated by cosmic-ray muons created in the Earth's upper atmosphere.  

The flux of atmospheric neutrinos has been a topic of study since the mid-1960's.  Early experiments~\cite{bib:Reines65,bib:Achar65} inferred the presence of atmospheric neutrinos by measuring the muon flux created by neutrino interactions taking place in rock surrounding a detector.  Subsequent studies of the atmospheric neutrino flux as a function of zenith angle~\cite{bib:Hirata92, bib:Fukuda99, bib:Ambrosio2000}, the ratio of electron and muon neutrinos~\cite{bib:imb, bib:frejus, bib:Hirata88, bib:Casper91, bib:Szendy}, and combined measurements~\cite{bib:Ashie05} have provided a more direct measurement of the atmospheric neutrino flux and revealed evidence for neutrino oscillations.  Results gathered from these experiments have been further verified by long baseline accelerator measurements~\cite{bib:Ahn03,bib:Michael06}, thereby providing strong constraints on the neutrino oscillation parameters.

The Sudbury Neutrino Observatory (SNO) is located in the Vale-Inco Creighton mine in Ontario, Canada at a depth of 2.092 km ($5890 \pm 94$ meters water equivalent) with a flat overburden\cite{bib:SNO00}.  The combination of large depth and flat overburden attenuates almost all cosmic-ray muons entering the detector at zenith angles less than \cosz = 0.4.  Because of this depth, SNO is sensitive to neutrino-induced through-going muons over a large range of zenith angles, including angles above the horizon.

This paper presents a measurement of the flux of muons traversing the SNO detector.  Measuring the through-going muon flux, as a function of zenith angle, for \cosz $< 0.4$ provides sensitivity to both the oscillated and un-oscillated portions of the atmospheric neutrino flux. Measuring the muon angular spectrum above this cutoff provides access to the flux of cosmic-ray muons created in the upper atmosphere.  This paper is divided as follows: Section~\ref{sec:detector} describes the experimental details of the SNO detector, Section~\ref{sec:flux} describes the Monte Carlo model used to predict the observed muon flux, Section~\ref{sec:data} describes the data collection and event reconstruction, and Section~\ref{sec:results} discusses the signal extraction and error analysis used for the measurements presented herein.

\section{The Sudbury Neutrino Observatory}
\label{sec:detector}

The Sudbury Neutrino Observatory is located at $46^\circ 28'30"$N latitude ($56^\circ33'$ magnetic north), $81^\circ12'04"$W longitude near the city of Sudbury, Ontario.   The center of the detector is at a depth of $2092 \pm  6$ meters from the Earth's surface. The Earth's surface immediately above the SNO detector is 309 meters above sea level.  Within a 5 km radius on the surface above the detector, the local topology lies between 300 and 320 meters above sea level, although small localizations with $\pm 50$ meters level variations do occur.  The norite rock that dominates the overburden is mostly oxygen (45\%), silicon (26\%), aluminum (9\%), and iron (4\%).  A combination of bore samples taken at different depths and gravity measurements taken at the surface show variations in the rock density, from 2.8 g/cm$^3$ near the surface and closer to 2.9 g/cm$^3$ in the vicinity of the detector.  A fault line located 70 meters southwest of the detector serves as a boundary to a deposit of granite/gabbro rock of similar density ($2.83 \pm 0.10$ g/cm$^3$) but slightly different chemical composition ($\langle Z^2/A \rangle$ of 5.84 versus 6.01).  An average rock density of $2.83 \pm 0.05$ g/cm$^3$ is used in overburden calculations, independent of depth.  The uncertainty in the density takes into account the variation as measured in the rock volume surrounding the detector. The total depth to the center of the SNO detector, taking into account air and water filled cavities, is $5890 \pm 94$ meters water equivalent.

The SNO detector itself includes a 600.5 cm radius acrylic vessel filled with 99.92\% isotopically pure heavy water (D$_2$O).  The 5.5-cm thick acrylic vessel is surrounded by 7.4 kilotons of ultra-pure H$_2$O encased within an approximately barrel-shaped cavity measuring 34 m in height and 22 m (maximum) in diameter.  A 17.8-meter diameter stainless steel geodesic structure surrounds the acrylic vessel.  The geodesic is equipped with 9456 20-cm photo-multiplier tubes (PMTs) pointed toward the center of the detector.   A non-imaging light concentrator is mounted on each PMT to increase the total effective photocathode coverage to 54\%. 

SNO is primarily designed to measure the solar neutrino flux originating from $^8$B decay in the sun above a threshold of several MeV by comparing the observed rates of the following three reactions:

\begin{eqnarray}
\begin{tabular}{llll}
$\nu_x + e^-$ & $\rightarrow$ & $\nu_x + e^-$ & (ES) \\
$\nu_e + d$ & $\rightarrow$ & $e^- + p + p$ & (CC) \\
$\nu_x + d$ & $\rightarrow$ & $e^- + p + n$ & (NC) 
\end{tabular}
\end{eqnarray}

The charged current (CC), neutral current (NC), and elastic scattering (ES) reactions outlined above are sensitive to different neutrino flavors.  Data taking in the SNO experiment is subdivided into three distinct phases, with each phase providing a unique tag for the final states of the neutral current interaction. In the first phase, the experiment ran with pure D$_2$O only.  The neutral current reaction was observed by detecting the 6.25-MeV $\gamma$-ray following the capture of the neutron by a deuteron.  For the second phase of data taking, approximately 0.2\% by weight of purified NaCl was added to the D$_2$O to enhance the sensitivity to neutrons via their capture on $^{35}$Cl.  In the third and final phase of the experiment, 40 discrete $^3$He or $^4$He-filled proportional tubes were inserted within the fiducial volume of the detector to enhance the capture cross-section and make an independent measurement of neutrons by observing their capture on $^3$He in the proportional counters.  Results from the measurements of the solar neutrino flux for these phases have been reported elsewhere~\cite{bib:SNO01, bib:SNO02a, bib:SNO02b, bib:SNO05, Aharmim:2008kc}.

Muons entering the detector produce Cherenkov light at an angle of 42$^\circ$ with respect to the direction of the muon track.  Cherenkov light and light from delta rays produced collinear to the muon track illuminate an average of 5500 PMTs.  The charge and timing distribution of the PMTs is recorded.  The amplitude and timing response of the PMTs is calibrated {\it in situ} using a light diffusing sphere illuminated by a laser at six distinct wavelengths~\cite{bib:Moffat05}.  This ``laser ball'' calibration is of particular relevance to the muon analysis since it provides a timing and charge calibration for the PMTs which accounts for multiple photon strikes on a single PMT.  Other calibration sources used in SNO are described in the references~\cite{bib:Dragowsky02, bib:SNO00}.

For a period at the end of the third phase of the experiment, a series of instrumented wire tracking chambers and scintillator panels were installed immediately above the SNO water cavity to provide a cross-check on the accuracy of the muon reconstruction algorithm.  Details of the apparatus and results obtained from this calibration are reported later in this paper.

\section{Simulation of Signal and Background Events}
\label{sec:flux}

Candidate neutrino-induced through-going muon events can arise from a variety of sources.  These include: (a) muons created from neutrino-induced interactions in the rock surrounding the SNO cavity; (b) muons created from neutrino-induced interactions in the H$_2$O volume surrounding the PMT support structure; (c) $\nu_\mu$ interactions that take place inside the fiducial volume but are misidentified as through-going muons; (d) $\nu_e$ interactions that take place inside the outlined fiducial volume but mis-reconstruct as through-going muons; (e) cosmic-ray muons created in the upper atmosphere that pass the zenith angle cut; and (f) events created by instrumental activity in the detector.  The first three event types are proportional to the $\nu_\mu$ atmospheric neutrino flux and can undergo oscillations. The $\nu_e$-induced flux is also proportional to the overall atmospheric neutrino flux, but the currently measured neutrino oscillation parameters indicate that their probability for undergoing oscillations is highly suppressed.  The last two entries constitute a genuine source of background to the signal.

\begin{figure}[!ht]
\includegraphics[width = 1.00\columnwidth,keepaspectratio=true,bb=0 0 525 375]{\thedirectory/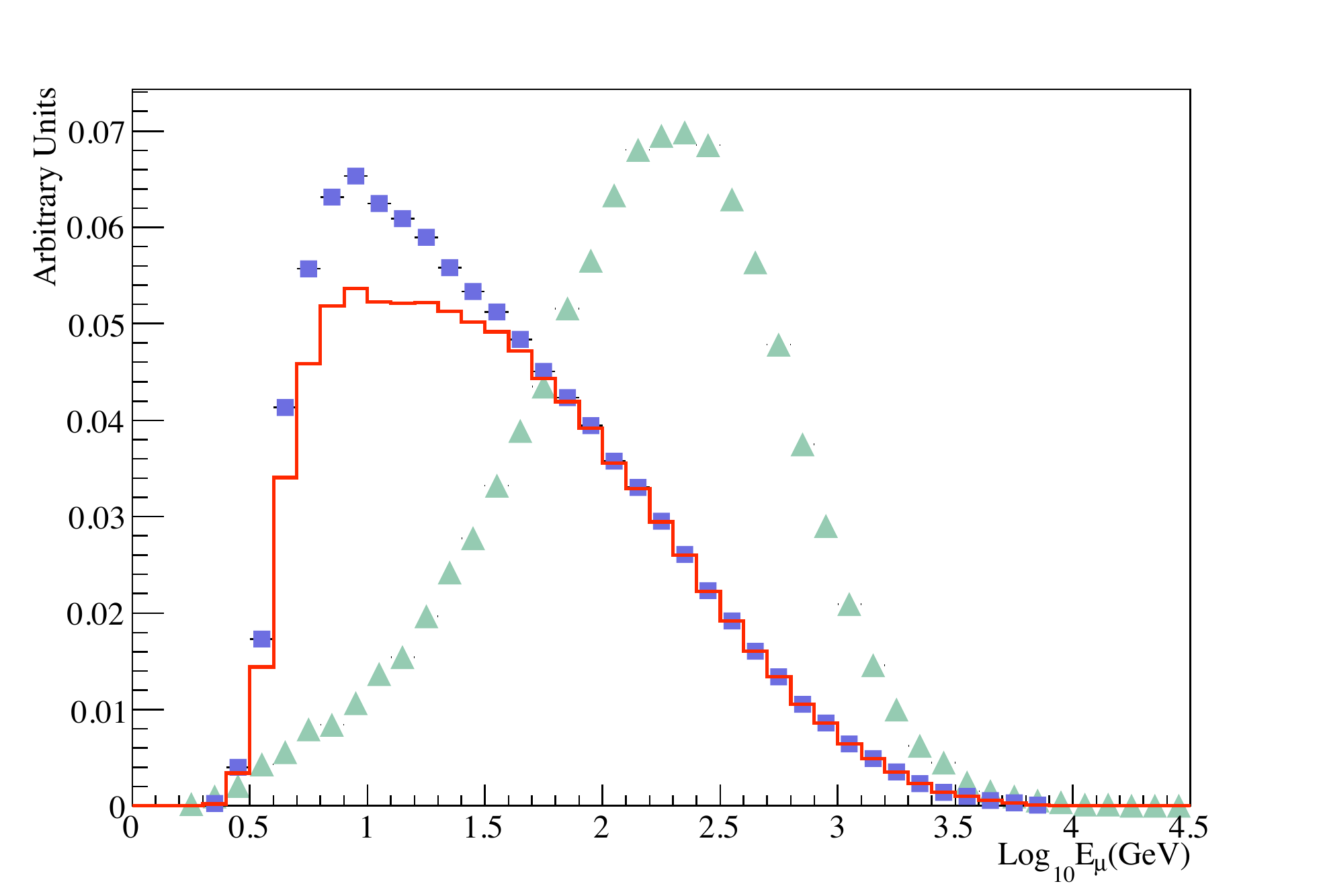}
\caption{The muon energy spectrum (given in log$_{10}(E_\mu)$) in  the SNO detector from cosmic-ray muons (triangles) as predicted from MUSIC, and from neutrino-induced muons (boxes) created in the surrounding rock as predicted from the Bartol 3D Monte Carlo.  The expected spectrum after oscillations is shown by the solid line. The distributions are not to scale.}
\label{fig:nuflux}
\end{figure}

In order to understand the measured neutrino-induced flux, a proper model of the initial neutrino flux and subsequent propagation is necessary.  SNO uses the Bartol group's three-dimensional calculation of the atmospheric neutrino flux~\cite{bib:Barr04}.  
Figure~\ref{fig:nuflux} shows the predicted flux for cosmic rays and muons from the interaction of muon neutrinos and anti-neutrinos as a function of muon energy.  The neutrino energy spectrum is correlated with the primary H and He cosmic flux, both of which are strongly constrained by data. The uncertainties that dominate the neutrino energy distribution relate to the primary cosmic-ray energy spectrum, the $\pi$ and $K$ production ratio, and hadronic cross sections.  Treatment of the systematic errors in the neutrino flux is discussed in greater detail in Ref~\cite{bib:Barr06} and~\cite{bib:Honda06}.  Current estimates of the neutrino flux uncertainties are approximately $\pm 15\%$ and depend strongly on neutrino energy. Because the normalization of the neutrino flux and the energy spectral shape are highly correlated, the fits to the data reported herein assume a fixed neutrino energy spectrum.  We also assume that the flux and energy spectra do not change significantly with solar activity. Although variations throughout the solar cycle are expected, the majority of this variation is confined to neutrinos of energy below 10 GeV, so the impact on the fluxes predicted at SNO is expected to be small.  A flux uncertainty of  $\pm 1\%$ is included to account for variations due to solar cycle activity.
  
Neutrino interactions in the rock surrounding the detector are simulated by the NUANCE v3 Monte Carlo neutrino event generator~\cite{bib:Capser02}. NUANCE includes a comprehensive model of neutrino cross sections applicable across a wide range of neutrino energies.  Neutrino quasi-elastic interactions are modeled according to the formalism of Llewellyn-Smith~\cite{bib:Smith}.  A relativistic Fermi gas model  by Smith and Moniz~\cite{bib:Moniz} is used to model the low momentum transfer effects in the nucleus.  The quasi-elastic cross-section depends strongly on the value of the axial mass used in the axial form factor.  Recent measurements from K2K~\cite{bib:K2K} and MiniBooNE~\cite{bib:MiniBooNE} show a higher value of the axial mass than previously reported ($m_{\rm axial} = 1.20 \pm 0.12$ GeV and $m_{\rm axial} = 1.23 \pm 0.20$, respectively).  Though this analysis uses the previous world average for the axial mass ($m_{\rm axial} = 1.03 \pm 0.15$ GeV)~\cite{bib:MA_old}, the systematic uncertainty encompasses these more recent measurements.  For the Fermi gas model, we assume a Fermi momentum of 225 MeV/c and a binding energy of 27 MeV for light elements such as oxygen, carbon, and silicon, and a 2.22 MeV binding energy for deuterium.  

For neutrino reactions where a single charged or neutral pion is resonantly produced, NUANCE employs a modified model of Rein and Sehgal~\cite{bib:Rein}.  Experimental constraints on this cross-section are of order $\pm 20\%$, not as strong as those placed on the quasi-elastic cross-section.  Many of the other parameters used for the quasi-elastic cross-section are also used for this process.  

The largest contributor to the atmospheric neutrino-induced muon flux is deep inelastic scattering of neutrinos in the surrounding rock where the hadronic invariant mass is above 2 GeV/c$^2$. The uncertainty on the cross-section for this process is strongly constrained by accelerator produced high energy neutrino experiments to $\pm 3\%$~\cite{bib:CCFR90, bib:CCFR84, bib:CDHSW87}.  The transition between resonance and deep inelastic scattering cross-section uses the methodology developed by Yang and Bodek~\cite{bib:Bodek}.
Other minor processes that can produce muons in the final states, such as coherent pion production and $\nu_\mu - e^-$ scattering, are also included.  

Transport of muons through the rock from neutrino-induced interactions  is calculated using the PROPMU muon transport code~\cite{bib:PROPMU}, which is integrated into the NUANCE Monte Carlo framework.  Rock compositions and densities consistent with measured values are implemented in PROPMU.  Simulation of muon transport in the D$_2$O and H$_2$O and subsequent detector response is handled by the SNO Monte Carlo and Analysis (SNOMAN) code.  SNOMAN propagates the primary particles and any secondary particles (such as Compton electrons) that are created, models the detection of the optical photons by the PMTs, and simulates the electronics response. The SNOMAN code has been benchmarked against calibration neutron, gamma, and electron data taken during the lifetime of SNO. With the exception of a few physics processes (such as optical photon propagation), widely used packages such as EGS4~\cite{bib:EGS4}, MCNP~\cite{bib:MCNP} and FLUKA~\cite{bib:FLUKA} are used in SNOMAN.  Explicit muon energy loss mechanisms such as ionization, pair production, bremsstrahlung, muon capture and decay, and photonuclear interactions are all included in the simulation,  allowing modeling of the muon track from rest energies up to several TeV. Energy losses due to photonuclear interactions are simulated using the formalism of Bezzukov and Bugaev~\cite{bib:Bezzukov73,bib:Bezzukov81}.  Production of secondary particles from muon interactions, which contributes to the total energy deposited in the detector, is included in the model as described above.

In addition to the through-going signal from atmospheric neutrinos, a number of backgrounds which have muon signatures are simulated in the analysis.  These include $\nu_\mu$ interactions inside the H$_2$O and D$_2$O volumes of the detector, and $\nu_e$ interactions that either have a muon in the final state or are misidentified as a through-going muon.  
Cosmic-ray muons incident on the detector constitute an additional source of high-energy muons and are treated separately.  Their flux is estimated using the formalism of Gaisser~\cite{bib:Gaisser},

\begin{widetext}
\begin{equation}
\label{eq:surface}
\frac{d^2\Phi_\mu}{dE_\mu d\Omega} \simeq I_0 \cdot E_\mu^{-\gamma} (\frac{1}{1+\frac{1.1E_\mu \cos{\theta_\mu}}{\rm 115~GeV}} + \frac{0.054}{1+\frac{1.1E_\mu \cos{\theta_\mu}}{\rm 850~GeV}}){\rm~~cm^{-2}~sr^{-1}~GeV^{-1}} 
\end{equation}
\end{widetext}

\noindent where $E_\mu$ and $\theta_\mu$ are the muon energy and zenith angle at the Earth's surface,  $\gamma \equiv 2.77 \pm 0.03 $ is the muon spectral index, and $I_0$ is a normalization constant. Although Eq.~\ref{eq:surface} is inaccurate at low energies, the minimum energy required for surface muons to reach the SNO detector is $\approx$ 3 TeV.  Transport of such high-energy muons in the rock is performed by the MUSIC muon transport code~\cite{bib:MUSIC}.   The average energy of these muons as they enter the SNO detector is $\approx$ 350 GeV.  After incorporating the detector response, simulated events for cosmic-ray and neutrino-induced muon candidates are used to construct probability distribution functions (PDFs). The reconstructed zenith angle is used to establish PDFs for both signal and background.  

\section{Event Selection and Reconstruction}
\label{sec:data}

\subsection{Livetime}

\livetimesection  Data were collected in discrete time intervals, or runs, that range from 30 minutes to 96 hours in length.  Runs that were flagged with unusual circumstances (presence of a calibration source in the detector, maintenance, etc.) were removed from the analysis.  The raw live time of the data set is calculated using a GPS-sychronized 10 MHz clock on a run-by-run basis, checked against an independent 50 MHz system clock, and corrected for time removed by certain data selection cuts.  The livetime of the dataset here is $\livetime \pm \livetimeerr$ days.  The livetime used in this analysis differs from previously published analyses because requirements that have a strong impact on solar neutrino analyses, such as radon activity levels, are relaxed here.  

\subsection{Event Reconstruction}

\begin{figure}[!ht]
\begin{center}
\includegraphics[width = 0.6\columnwidth,keepaspectratio=true,bb=0 0 525 525]{\thedirectory/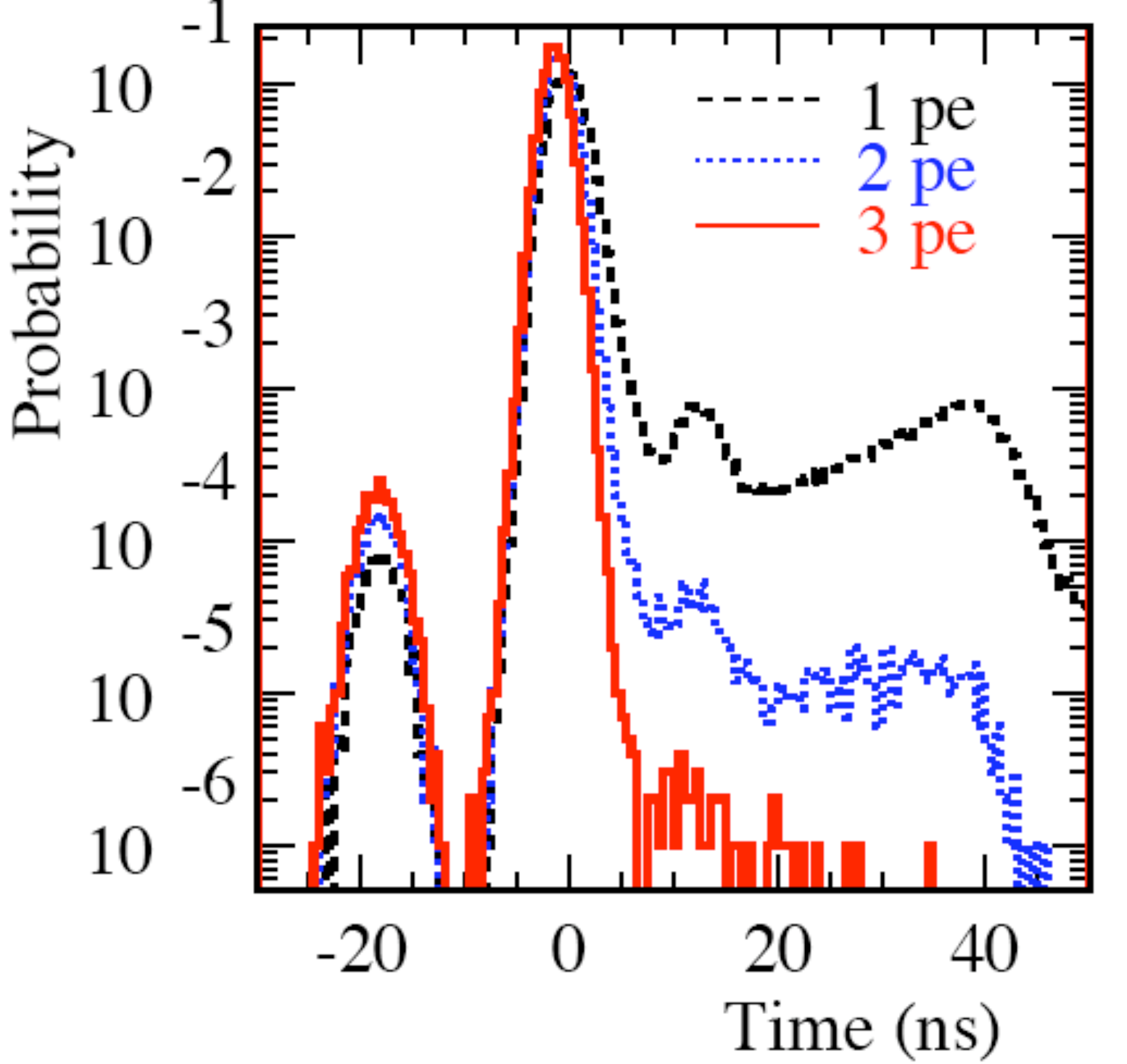}
\caption{The probability distribution of PMT firing times based on simulation for events with one (dashed), two (dotted), and three (solid) photons striking the photocathode.}
\label{fig:timing}
\end{center}
\end{figure}

The SNO detector has a nearly ideal spherically symmetric fiducial volume.  The algorithm used to reconstruct muon candidates makes use of this symmetry in finding the best fit track.  A two-tiered algorithm is used whereby a preliminary track is reconstructed which later serves as a seed for a more comprehensive fit to the muon candidate event.   In the preliminary fit, the entrance position is determined by looking at the earliest hit PMTs, and the exit position is determined by the charge-weighted position of all fired PMTs. In our spherical geometry, the impact parameter is the distance from the center of the sphere to the midpoint of the line connecting the
entrance and exit points. The fitter corrects the track fit for biases in charge collection and geometry, and provides a first estimate of the direction of the incoming muon. The first-order reconstructed track is then passed to a full likelihood fit to determine the muon track parameters to greater accuracy. The likelihood fit uses three distributions: (a) the number of detected photoelectrons, (b) the PMT charge distribution, and (c) the PMT timing distribution.  The charge and timing distributions are conditional on the number of photoelectrons incident on a given PMT.  These distributions are corrected according to biases measured during laser calibrations.  Figure~\ref{fig:timing} shows the timing distribution expected for different numbers of photoelectrons that are above detection threshold.  Both changes in pre and post-pulsing and biasing can be seen in the timing distributions due to multiple photon hits on a given PMT event.

\begin{figure*}[!ht]
\includegraphics[width = 2.00\columnwidth,keepaspectratio=true,bb=0 0 525 375]{\thedirectory/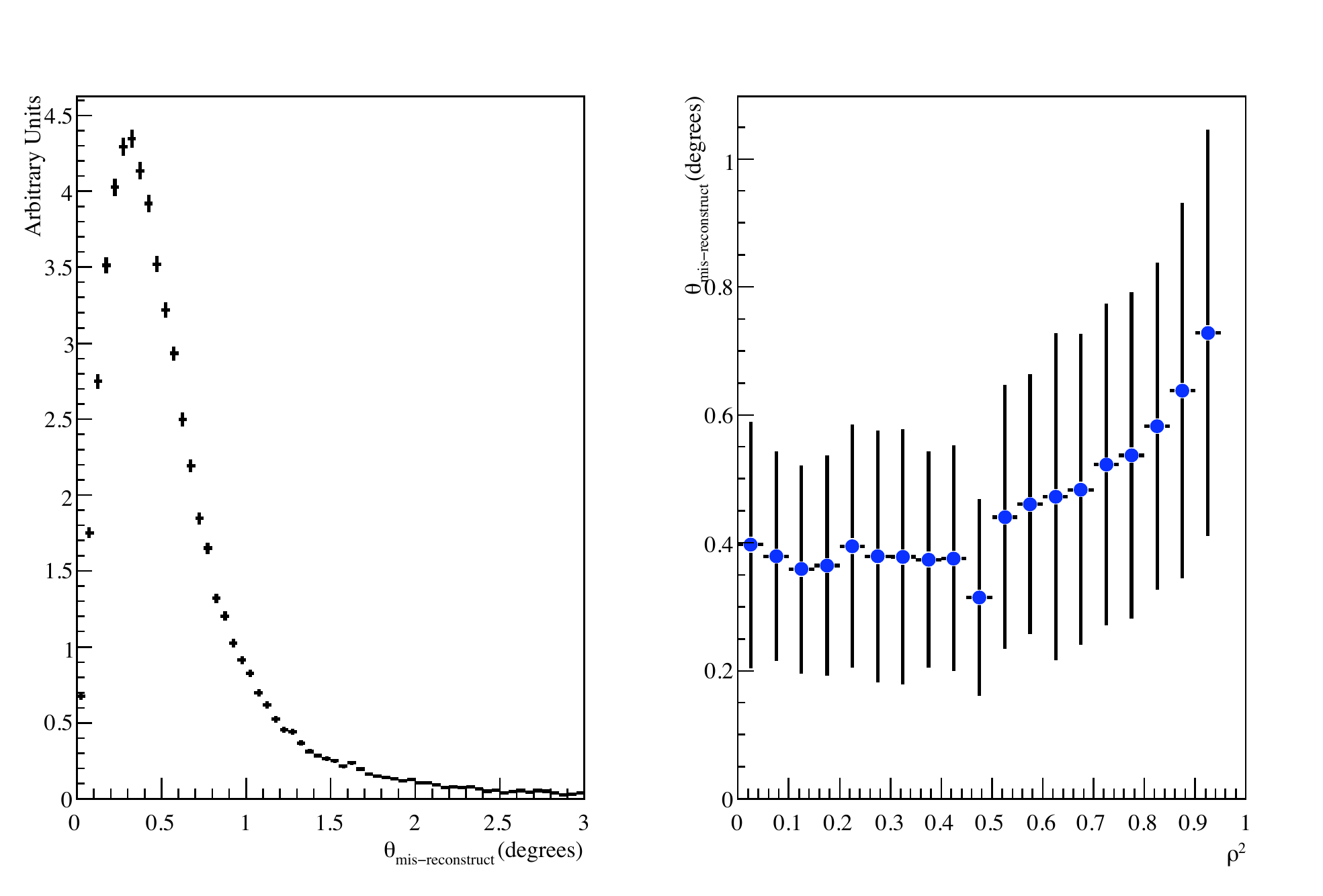}
\caption{Distribution of mis-reconstructed zenith angle (left) and as a function of normalized impact parameter, $\rho^2 = b_\mu^2 / (850~\rm{cm})^2$, (right) for Monte Carlo cosmic-ray muons.  The points and error bars on the right-hand side plot refer to the mean and RMS of the mis-reconstructed angle, respectively.}
\label{fig:fit_angle}
\end{figure*}

\begin{figure*}[!ht]
\includegraphics[width = 2.00\columnwidth,keepaspectratio=true,bb=0 0 525 375]{\thedirectory/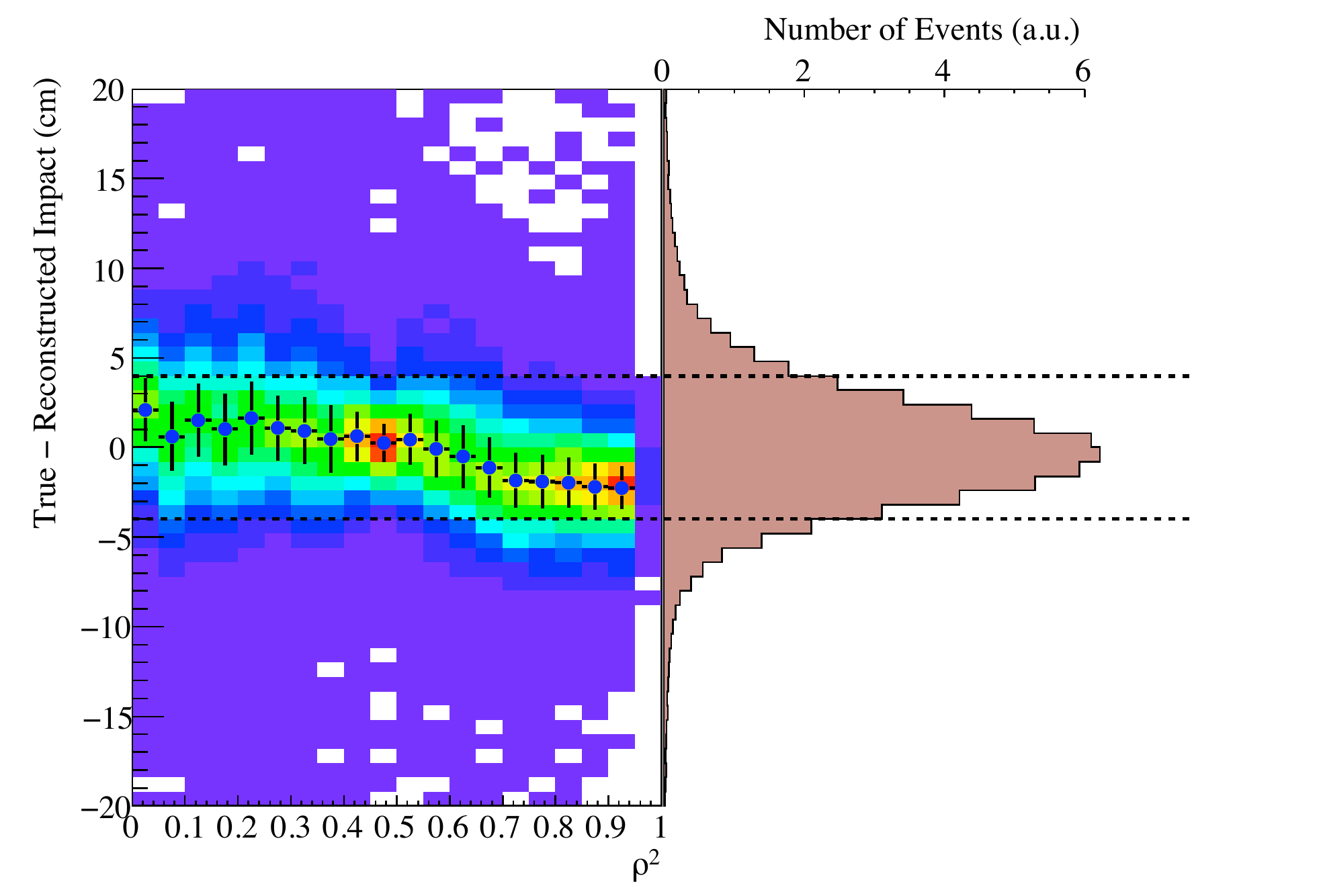}
\caption{Left: Comparison of true and reconstructed impact parameter versus normalized impact parameter, $\rho^2 = b^2_\mu / (850~\rm{cm})^2$ for Monte Carlo cosmic-ray muons.  Data points indicate mean and error bars for a given impact parameter. Right: Projection of difference between reconstructed and generated tracks.  Dashed lines indicate uncertainty in impact parameter reconstruction as adopted for this analysis.}
\label{fig:fit_impact}
\end{figure*}

The use of conditional distributions helps remove reconstruction biases due to multiple photoelectrons detected on a single PMT.  This is important for impact parameter values close to the PMT support structure ($b_\mu \approx 830$ cm).  The quality of reconstruction for the bias of the fitted track and for the mis-reconstruction angle as a function of impact parameter were examined. Figure~\ref{fig:fit_angle} shows the cosine of the mis-reconstruction angle, defined as the dot-product between the true muon direction vector and the reconstructed vector.  Approximately 87\% (97\%) of all simulated muons with an impact parameter of less than 830 cm reconstruct within 1$^\circ$ (2$^\circ$) of the true track direction; respectively (see Figure~\ref{fig:fit_angle}).   Monte Carlo studies also show that bias effects on the reconstructed impact parameter to be less than $\pm 4$ cm (see Figure~\ref{fig:fit_impact}).

\begin{figure}[!ht]
\includegraphics[width = 1.00\columnwidth,keepaspectratio=true,bb=0 0 525 375]{\thedirectory/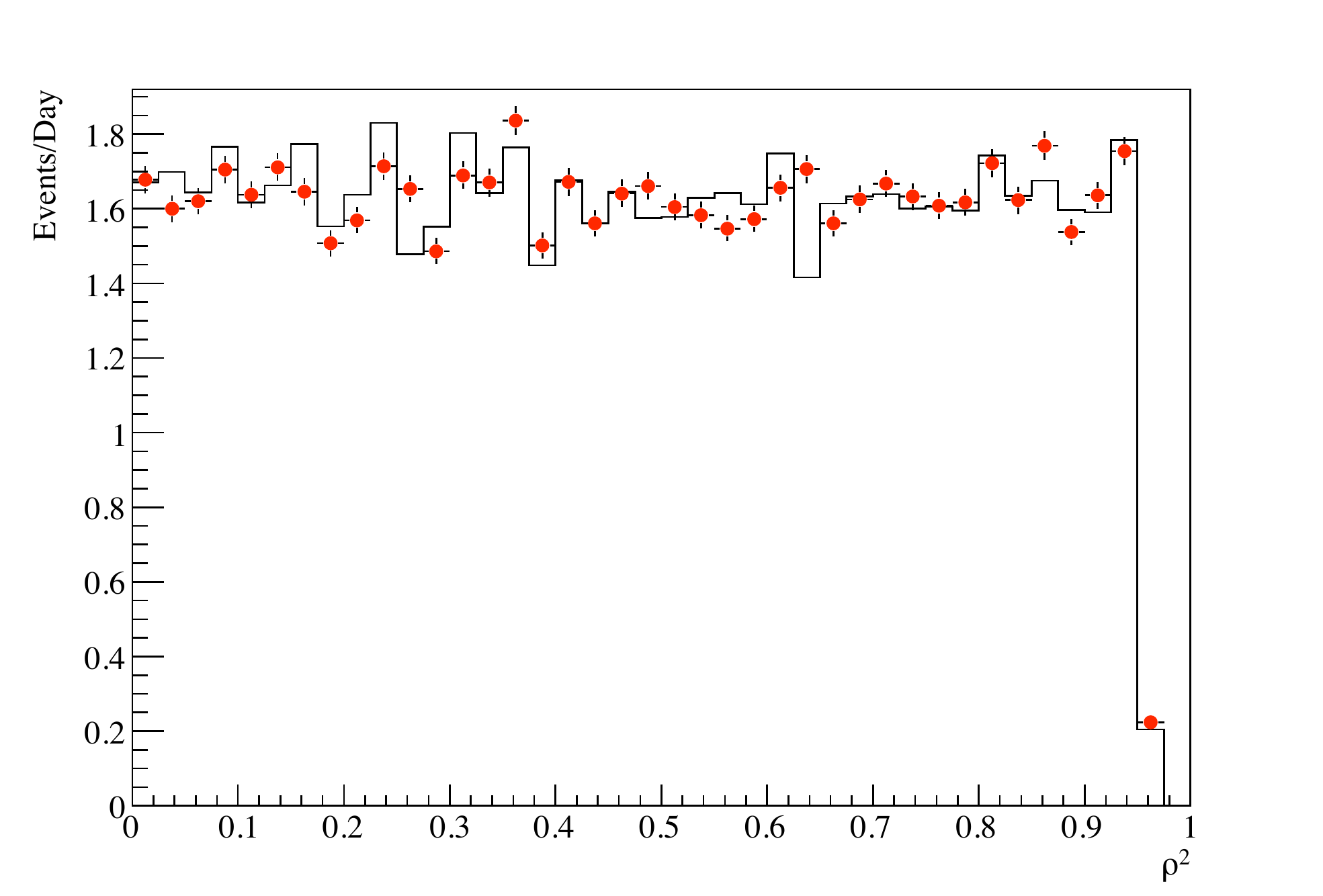}
\caption{Distribution of the normalized impact parameter, $\rho^2 = b_\mu^2 / (850~\rm{cm})^2$ for cosmic-ray muon data (points) and Monte Carlo (solid line).}
\label{fig:rho}
\end{figure}

\subsection{Event Selection}

After run selection, low-level cuts are applied to measurements of PMT outputs before reconstruction in order to separate through-going muon candidate events from instrumental background activity.  We require a minimum of 500 valid (or calibrated) PMT hits for an event to be a muon candidate.  Events with more than 250 hit PMTs within a 5 $\mu$s window of a previously tagged event, or when 4 or more such events occur within a 2 s window, are identified as burst events.  Burst events are often associated with instrumental backgrounds and are removed from the analysis.  Instrumental activity typically has broad PMT timing distributions and/or low total charge; events with these characteristics are removed. Finally, events that possess 4 or more hit PMTs in the aperture of the D$_2$O vessel (neck) are removed from the data to eliminate occurrences where light enters the detector from the top of the acrylic vessel.  

\begin{figure}[!t]
\includegraphics[width = 1.0\columnwidth,keepaspectratio=true,bb=50 80 705 780]{\thedirectory/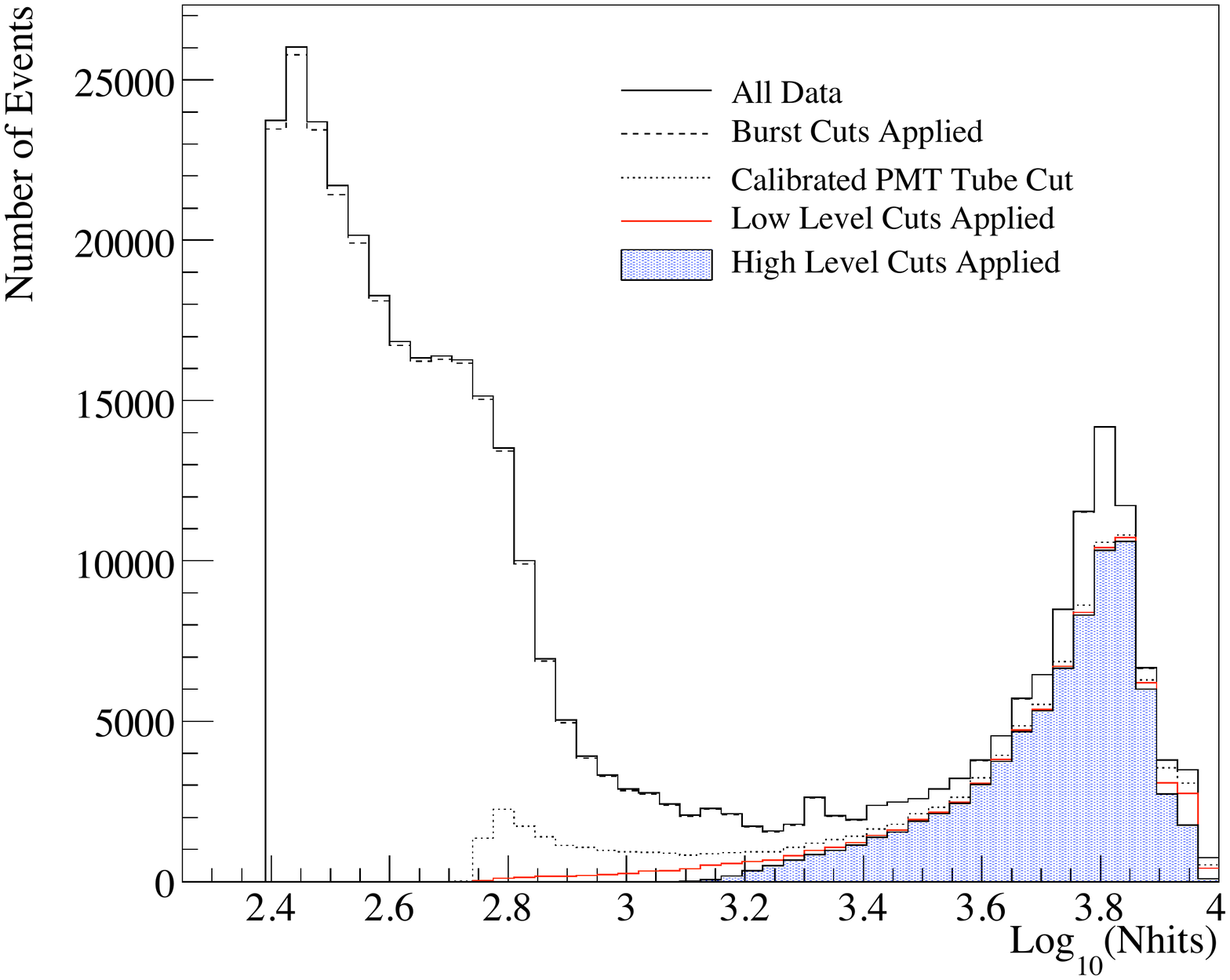}
\caption{The number of muon candidate events as a function of the log of the number of tubes that fire.  Plot shows events with no cuts applied (solid), after the burst cut (dashed), calibrated tubes cut (dotted), all low-level cuts (red) and high-level cuts (filled area) are applied.}
\label{fig:cuts}
\end{figure}

A series of high-level analysis cuts use reconstructed track parameters to isolate a pure through-going muon data set. A cut on the reconstructed impact parameter of $b_\mu < 830$ cm is applied to the data to ensure accurate reconstruction of through-going tracks.  These cuts define a total fiducial area of 216.42 m$^2$ and a minimal track length of 367 cm. The minimum (mean) muon energy needed to traverse this length of track is 0.8 (2.6) GeV.  Muon events characteristically produce ample amounts of light in the detector.  The number of Cherenkov photons produced by the muon, scaled by the appropriate detection efficiency for photons produced at the given impact parameter, is reconstructed for each candidate event. Each track is required to possess a minimum of 2000 detected photoelectrons. A cut is also made on the estimated energy loss ($dE/dX$) of the muon. The quantity $dE/dX$ is determined from the amount of detected light, corrected for geometric and photon attenuation effects, divided by the reconstructed track length.  The $dE/dX$ variable depends on both the ionization and radiation losses, and has a peak at around 225 MeV/m.  Events with $dE/dX \ge 200$ MeV/m are retained for further analysis.  Further, a cut is imposed on the fraction of photoelectrons within the predicted Cherenkov cone for the muon track, and on the timing of these in-cone photons. Finally, a linear combination (Fisher discriminant) formed from the fraction of in-time hits and the time residuals from the muon fit is used to reduce the contamination of contained atmospheric neutrino events in our final data sample. A list of all cuts and their effects on the data is shown in Table~\ref{tab:cuts} and in Figure~\ref{fig:cuts}.  

%
%
\begin{table}[htdp]
\caption{Summary of low- and high-level cleaning cuts applied to the data and their effect on the data population.  Cuts are applied in sequence as they appear in this table.}
\begin{center}
\begin{tabular}{|c|c|c|}
\hline
Level & Type of Cut & No. of Events \\
\hline
	& Raw number of tubes firing $>$ 250 & 378219 \\
	& Timing and burst requirements & 375374 \\
Low	& Number of calibrated tubes firing & 100396 \\
	& Raw PMT charge requirement & 85703 \\
	& Raw PMT timing RMS & 84414 \\
	& Number of neck tubes firing & 84038 \\
\hline
	& Impact parameter $\le 830$ cm & 80165 \\
	& Fit number of photoelectrons & 79998 \\
High	& Energy loss ($dE/dX$) & 79268 \\
	& Linear discriminant cut & 77321 \\
	& Cherenkov cone in-time fraction& 77321 \\
	& Cherenkov cone fraction of tubes firing  & 77263 \\
\hline
Zenith & \cosz$ > 0.4$ & \downwardevents \\
            & \cosz$ < 0.4$ & \upwardevents \\
\hline	
\end{tabular}
\end{center}
\label{tab:cuts}
\end{table}%

The reconstructed cosmic-ray tracks, after all selection criteria are applied, exhibit a flat distribution versus impact area, as expected (see Figures~\ref{fig:fit_impact}~and~\ref{fig:rho}). The reconstruction efficiency is also robust to a number of changes in the optical and energy loss model of the reconstruction.  Monte Carlo simulation shows that changes in the parameters of the detector model, including Rayleigh scattering, secondary electron production, PMT photocathode efficiency, and PMT angular response, all have minimal impact on the reconstruction performance.  An uncertainty on the reconstruction efficiency of $\pm 0.3\%$ is assigned due to detector model dependence.  The efficiency of the event selection depends most sensitively on the energy loss parameter, $dE/dX$.  We determine the energy loss model uncertainty on the reconstruction efficiency by studying the level of data-Monte Carlo agreement.  For through-going cosmic-ray muons it is $\pm 0.2\%$ and for neutrino-induced muons it is $\pm 2.5\%$.   Similar uncertainties in reconstruction efficiency arise from the PMT charge model invoked in reconstructing events and in the rejection of events from the linear discriminat cut previously mentioned.  This leads to a $\pm 0.05\%$ ($\pm 1.0\%$) and $\pm 0.37\%$ ($\pm 2.1\%$) uncertainty on the cosmic-ray (neutrino-induced) flux from the charge and linear discriminant cuts, respectively.  The differences seen in the two muon sources are due to the differences in muon energy distribution.  Monte Carlo studies of cosmic-ray events that pass through the SNO detector show the total event selection cut efficiency to be \downwardefficiency for through-going muons.

\subsection{Quality Checks and Calibration}

Neutrino-induced muons have a minimum energy of about 2 GeV with significant intensity extending into the hundreds of GeV range.  There is no readily available controlled calibration source that can provide multi-GeV muons as a benchmark to test the reconstruction algorithms.  Instead, a number of checks have been carried out to test the performance of the Monte Carlo by comparing with data.  

\begin{figure}[!ht]
{
    \includegraphics[width=0.95\columnwidth,keepaspectratio=true,bb=0 0 525 375]{\thedirectory/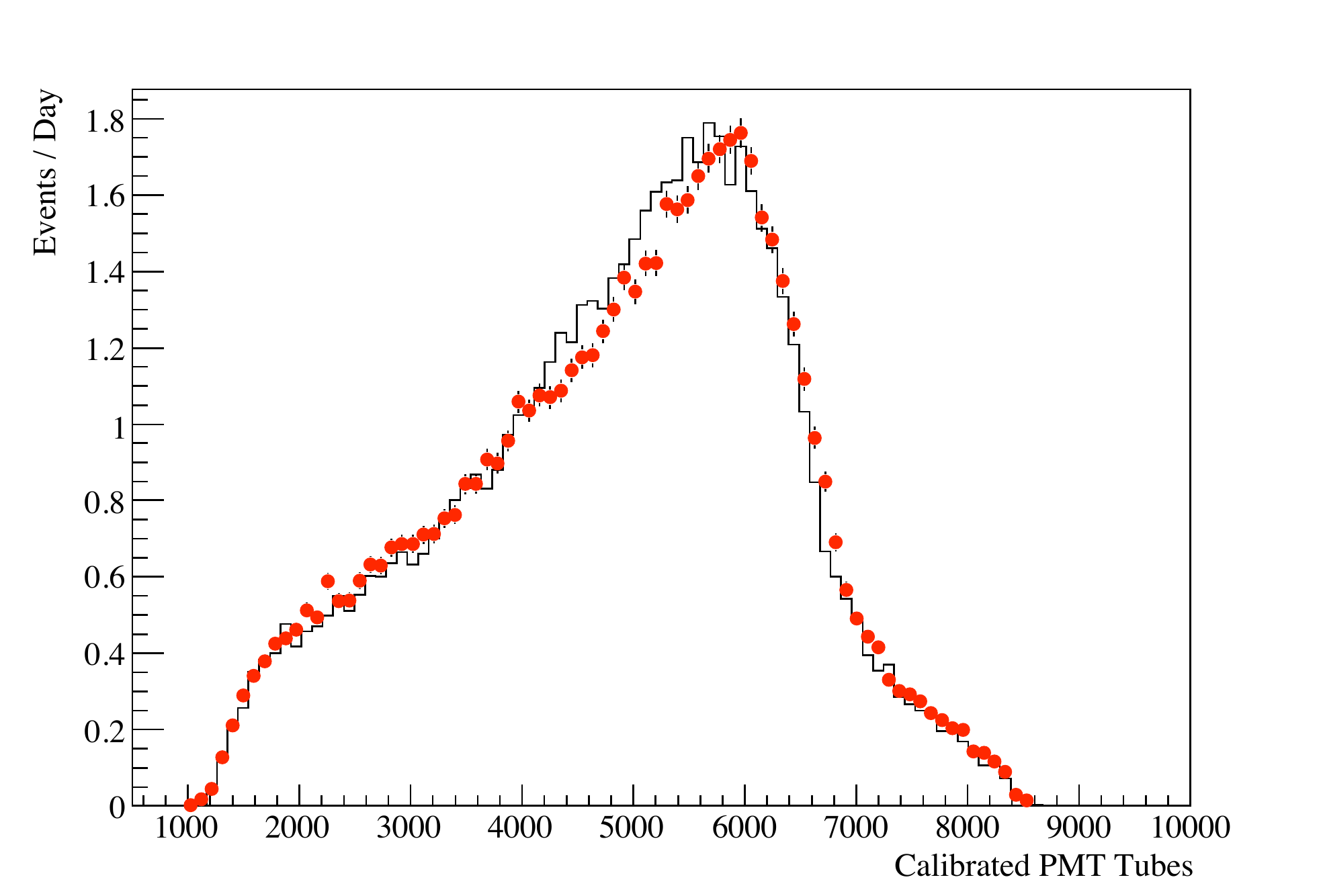}
}
\caption{The distribution of hit tubes (NHit) for muon events that pass all cuts for data (crosses) and Monte Carlo (line).  The Monte Carlo has been normalized to the total number of events seen in the data.  Only statistical errors are shown in the figure.}
\label{fig:dedx_nhit}
\end{figure}

The majority of checks are performed using muons that reconstruct in the downward direction ($\cos{(\theta_{\rm zenith})} > 0.4$).  Although the total energy of these cosmic-ray muons extends much higher than those from neutrino-induced muons, the amount of energy deposited in the detector by both is very similar.  The slight differences between the two energy loss distributions arise mainly from the relative fraction of radiative processes contributing to the energy loss mechanisms.  Figure~\ref{fig:dedx_nhit} shows the distribution of total hit PMTs (NHit) for reconstructed cosmic-ray muons. In general, there is good agreement between data and Monte Carlo simulations.  


The neutrino oscillation analysis is particularly sensitive to two parameters: (a) the fitter bias in reconstructing events at the edge of the impact parameter acceptance and (b) the angular resolution of the muon zenith angle.  The former affects the fiducial area of the experiment, while the latter affects the neutrino angular distribution, thereby affecting the oscillation parameter extraction. To test the accuracy of the muon track reconstruction, data and Monte Carlo distributions for cosmic-ray muons are compared at high impact parameters.  A chi-square ($\Delta \chi^2_\nu$) test is performed between data and Monte Carlo simulations for impact parameter distributions under different models of impact parameter bias and resolution:  (a) a constant shift ($b_\mu' = b_\mu + \delta x$), (b), a linear bias ($b_\mu' = b_\mu \cdot (1 + \delta x)$), and (c) a larger impact parameter resolution.  Results are summarized in Table~\ref{tab:KS} and show that our reconstruction model is consistent with the data and a small reconstruction bias ($\delta b_\mu < 4$ cm at 68\% C.L.).

\begin{figure}[!t]
\includegraphics[width = 0.95\columnwidth,keepaspectratio=true,bb=25 25 525 525]{\thedirectory/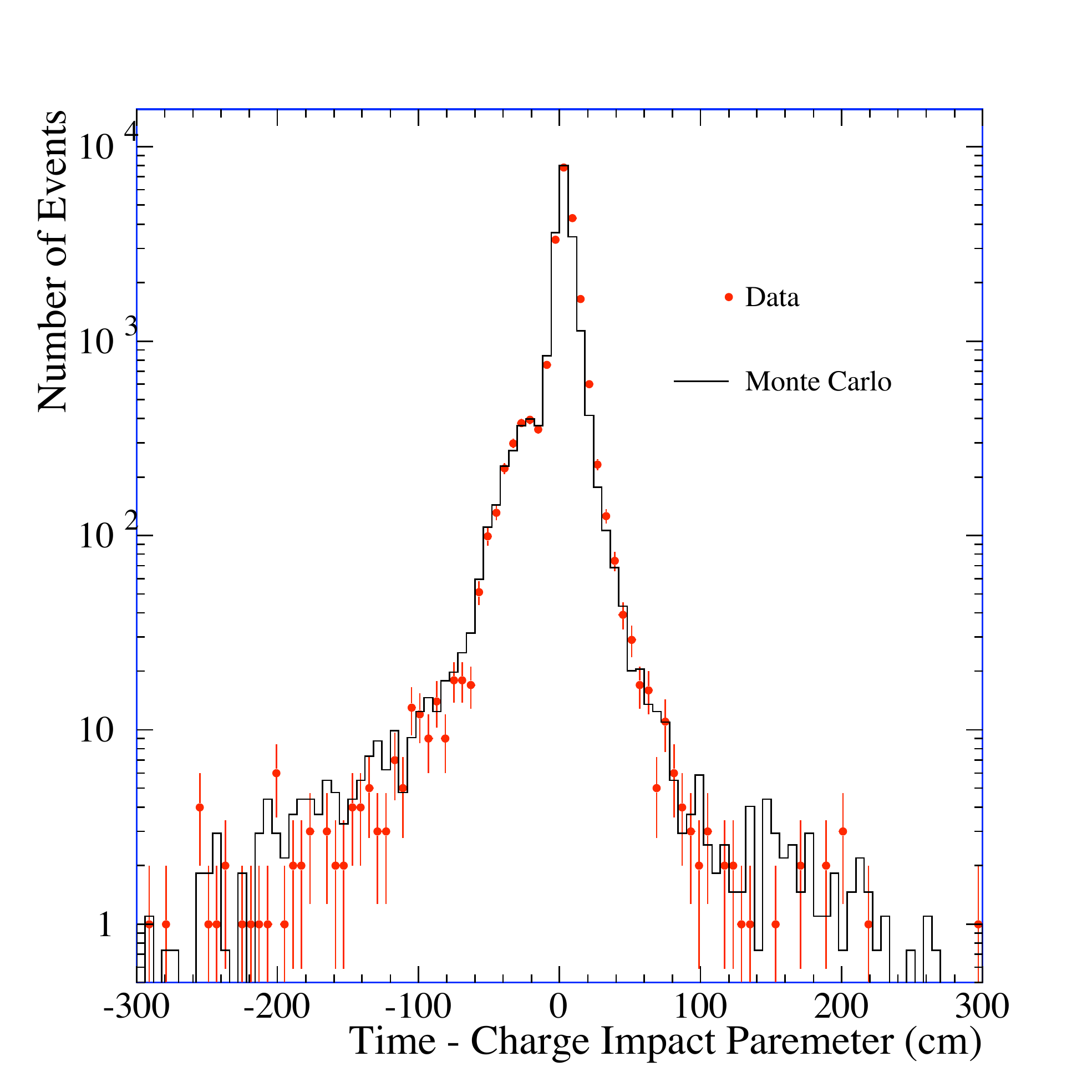}
\caption{Difference in reconstructed impact parameter $(b_\mu)$ using ``time-only" and ``charge-only" information in the likelihood minimization scheme for muons that pass all analysis cuts (points).  The mean difference between the two methods shows a 1.1 cm offset in the reconstruction of the impact parameter in comparison to simulations (solid line).}
\label{fig:time_charge}
\end{figure}

The muon fitter uses both time and charge to reconstruct muon tracks.  To test the robustness of the algorithm, tracks are fit under two conditions, using ``charge only"  and``time-only" information in order to search for potential biases in reconstruction (Figure~\ref{fig:time_charge}).  Differences between the two tracking methods are well-modeled by the Monte Carlo simulations(see also Table~\ref{tab:KS}).  The observed 1.1 cm shift is interpreted as the lower limit on the accuracy of the muon impact parameter reconstruction.

A more comprehensive test of the muon tracking algorithm is to compare tracks reconstructed in the SNO detector with an external charged particle tracking system.  A muon tracker was installed immediately above the SNO detector. The apparatus took data for a period at the end of the third phase of the SNO experiment.  A total of four wire planes, each spanning an area of approximately $2.5 \times 2.4$ m$^2$ and containing 32 instrumented wire cells, was arranged in alternating orthogonal coordinates to provide two dimensional track reconstruction. The overall structure was tilted at a 54-degree angle due to pre-existing space constraints.  Three large scintillator panels spanned the area covered by the wire chambers, and provided the trigger for the wire chamber readout.  A common trigger was also sent to the main SNO data acquisition system to synchronize events.  A total of 94.6 days of live time was recorded by the muon calibration unit. Track reconstruction of muon candidates from this instrument were compared with the SNO tracking algorithm. Further details on this calibration system will be described in a future article.

High-energy muons passing through this external muon tracking system and the SNO detector were reconstructed by both systems, providing a calibration check for the main SNO tracking algorithm.  A total of 30 tracks were used for comparisons of track reconstruction between both systems.  This test confirms the accuracy of the muon track reconstruction to better than $0.62^\circ \pm 0.12^\circ$.  A small shift in the reconstructed impact parameter is observed in the data, consistent with the limits from the previously mentioned tests. We take the uncertainty on the impact parameter reconstruction as $\pm 4$ cm.

\begin{table}[htdp]
\caption{List of consistency checks for the through-going muon analysis, including $\Delta \chi^2_\nu$ tests on the radial distribution of cosmic-ray muons, charge and time reconstruction differences, and the external muon tracking data.  See text for more details.}
\begin{center}
\begin{tabular}{|l|c|}
\hline \hline
Method & Bias \\
\hline \hline
\multicolumn{2}{|l|}{\bf $\Delta \chi^2_\nu$ Tests}\\
\hline
Impact Parameter Bias & $ ^{+1.0\%}_{-0.6\%}$\\
Impact Parameter Shift & $ ^{+3.8}_{-2.5}$ cm \\
Impact Parameter Resolution & $+ 8.5$ cm \\
\hline
\multicolumn{2}{|l|}{\bf Charge-Time Reconstruction}\\
\hline
Impact Parameter Shift & $\pm 1.1$ cm\\
\hline
\multicolumn{2}{|l|}{\bf External Muon Chambers} \\
\hline
Angular Resolution & $< 0.6^\circ$\\
Impact Parameter Bias & $4.2 \pm 3.7$ cm \\
\hline \hline
\end{tabular}
\end{center}
\label{tab:KS}
\end{table}%

A final check is performed on the time interval distribution between muon events. A fit to an exponential function yields an average time constant of $\sim 21$ minutes and a $\chi^2/$dof of 107.7/98, consistent with the hypothesis of a random arrival time of cosmic-ray muons, as expected.

\subsection{Expected Neutrino Signal and Background Rates}

Neutrino-induced muons from the H$_2$O and rock surrounding the SNO detector were simulated in the manner outlined in the previous sections.  A zenith angle cut of \cosz $\le 0.4$ was imposed to reject cosmic-ray muons from our neutrino-induced signal.  Under the assumptions of the Bartol~\cite{bib:Barr04} atmospheric neutrino flux and no oscillations, SNO expects a total of \signalrate neutrino-induced events per year passing all cuts.  
A full breakdown of the expected signal contribution is shown in Table~\ref{tab:backgrounds}.

The efficiency for reconstructing these signal events is not as high as that for primary cosmic-ray muons because some of neutrino-induced events stop within the detector volume.  The total efficiency is defined as the ratio between the number of through-going cosmic rays that reconstruct with an impact parameter less than 830 cm that pass all cuts versus the number of through-going muon events with a generated impact parameter less than 830 cm.  If the muon is genuinely through-going (exits the fiducial area of the detector), the total efficiency is \upwardefficiency, based on Monte Carlo studies.

%
%


\begin{table*}[!ht]
\caption{Summary of Monte Carlo expected signal and background rates contributing to the neutrino-induced muon analysis, after all cuts, for the full zenith angle range of $-1<$\cosz$<0.4$ and the unoscillated region of $0<$\cosz$<0.4$.  Errors include full systematic uncertainties assuming no correlations (see Table~\ref{tab:systematics} for more details). Neutrino induced interaction rates assume no oscillations.  The last entry in the table shows the measured muon rate passing all cuts.}
\begin{center}
\begin{tabular}{|c|c|c|}
\hline 
Source & \multicolumn{2}{|c|}{Rate (yr$^{-1}$)}\\
\hline
Zenith Range & ~~$-1 <$ \cosz $ < 0.4$~~ & ~~$0 < $ \cosz $< 0.4$~~ \\
\hline \hline
Through-going $\nu_\mu$ rock  interactions & \rockrate & \hrockrate \\
Through-going $\nu_\mu$ water interactions & \waterrate & \hwaterrate \\
Internal $\nu_\mu$ interactions & \internalrate & \hinternalrate \\
Internal $\nu_e$ interactions & \nuerate & \hnuerate \\
{\bf Total Signal} & \signalrate & \hsignalrate \\
\hline
Cosmic ray $\mu$ & \cosmicrate & \hcosmicrate \\
Instrumental contamination &\instrumentalrate & \hinstrumentalrate \\
{\bf Total Background} & \backgroundrate & \hbackgroundrate \\
\hline
{\bf Total Expected Rate} & \totalrate & \htotalrate \\
\hline
{\bf Detected Rate} & \dataupward &  \datahupward \\
\hline \hline
\end{tabular}
\end{center}
\label{tab:backgrounds}
\end{table*}%
 
SNO also has a small acceptance for neutrino-induced muons whose interaction vertex resides inside a fiducial volume defined by the 830-cm radius.  Most of these events are removed by the energy loss cut.  From Monte Carlo studies, contamination of \internalrate contained $\nu_\mu$-events per year is expected in the data.  As these events also depend on the flux and neutrino oscillation parameters, they are included as part of the final signal extraction.  A small number of internal neutrino events also come from $\nu_e$ interactions which reconstruct as through-going muons. The rate of these events is \nuerate events/year.  The cosmic-ray muon background passing all cuts is estimated to be \cosmicrate events per year.  Finally, a negligible amount of instrumental backgrounds are expected to contaminate the muon signal.  The majority of such instrumentals are due to burst activity present in the detector. A bifurcated analysis comparing the high-level cuts against the low-level cuts is performed so as to  determine the amount of contamination of these instrumental events in our data~\cite{bib:Vadim,bib:Bergbusch}.   In addition, events that are explicitly tagged as burst events are used to test the cut effectiveness in removing instrumental contamination.  Both tests predict an instrumental background contamination rate of \instrumentalrate events per year.  

\section{Flux and Oscillation Results}
\label{sec:results}

\subsection{Cosmic-Ray Muon Flux}

In order to minimize the possibility of introducing biases, a two-tier blind analysis procedure is employed.  First, only a fraction ($ \simeq 40\%$) of the data was open for analysis. Second, a fraction of muon events was removed from the data set using a zenith angle-dependent weighting function unknown to the analyzers.  Only after all fitter and error analyses were completed were both blindness veils lifted.

A total of \downwardevents muon candidates passing all selection cuts are reconstructed with a zenith angle of  $0.4 < $\cosz$ < 1$ for the \livetime-day dataset.  The data collected corresponds to an exposure of \luminosity.  The total measured cosmic-ray muon flux at SNO, after correcting for acceptance, is \flux, or  \dailyrate muons/day passing through a 830-cm radius circular fiducial area.  

One can define the vertical muon intensity per solid angle $I^v_\mu$ by the expression:

\begin{equation}
\label{eq:vertical}
I^v_\mu \cos{(\theta_{z,i}}) = \frac{1}{L \cdot \Omega_i \cdot \epsilon \cdot A} \sum_{j=1}^{N_i} \cos{\theta_{z,j}}
\end{equation}

\noindent where $N_i$ is the number of events in a given solid angle bin $\Omega_i$ and zenith angle $\theta_{z,i}$, $L$ is the livetime of the measurement, $\epsilon$ is the detection efficiency for through-going muons, and $A$ is the fiducial area.  Given the flat overburden, it is possible to express Equation~\ref{eq:vertical} in terms of the slant depth, $x_{\rm SNO}$.  To compare to other vertical flux measurements, SNO rock can be corrected to standard rock, CaCO$_3$, using the relation:

\begin{equation}
x_{\rm std} = 1.015 x_{\rm SNO} + \frac{x_{\rm SNO}^2}{4 \times 10^5 {\rm m.w.e}}
\end{equation}

\noindent where $x_{\rm SNO}$ is the slant depth expressed in meters water equivalent.  There exists an additional $\pm 1\%$ model uncertainty in converting from SNO to standard rock which is estimated from differences that arise between the MUSIC and PROPMU energy loss models. Flux values for slant depths ranging from 6 to 15 km water equivalent are presented in Table~\ref{tab:vmuon}. 

%
%
\begin{table*}[htdp]
\caption{Intensity for standard (CaCO$_3$) rock as a function of slant depth (in meters water equivalent) for muons passing all cuts and which reconstruct with \cosz $> 0.4$.  Only statistical errors are shown.}
\begin{center}
\begin{tabular}{|c|c|c|}
\hline
Slant depth & Events & Intensity (standard rock)  \\
(meters w.e.) &  &  (cm$^{-2}$ s$^{-1}$ sr$^{-1}$)\\ 
\hline
6225 & 4203 & $ (  3.71  \pm 0.53) \times 10^{-10} $ \\ 
6275 & 3905 & $ (  3.47  \pm  0.50) \times 10^{-10} $ \\ 
6325 & 3576 & $ (  3.20  \pm 0.46) \times 10^{-10} $ \\ 
6375 & 3371 & $ (  3.05  \pm 0.44) \times 10^{-10} $ \\ 
6425 & 3238 & $ (  2.95  \pm 0.43) \times 10^{-10} $ \\ 
6475 & 3000 & $ (  2.75  \pm  0.40) \times 10^{-10} $ \\ 
6525 & 2737 & $ (  2.53  \pm 0.37) \times 10^{-10} $ \\ 
6575 & 2598 & $ (  2.42  \pm 0.36) \times 10^{-10} $ \\ 
6625 & 2369 & $ (  2.23  \pm 0.33) \times 10^{-10} $ \\ 
6675 & 2182 & $ (  2.07  \pm 0.31) \times 10^{-10} $ \\ 
6725 & 2038 & $ (  1.94  \pm 0.29) \times 10^{-10} $ \\ 
6775 & 1911 & $ (  1.84  \pm 0.28) \times 10^{-10} $ \\ 
6825 & 1831 & $ (  1.77  \pm 0.27) \times 10^{-10} $ \\ 
6875 & 1668 & $ (  1.63  \pm 0.25) \times 10^{-10} $ \\ 
6925 & 1552 & $ (  1.52  \pm 0.24) \times 10^{-10} $ \\ 
6975 & 1377 & $ (  1.36  \pm 0.21) \times 10^{-10} $ \\ 
7025 & 1359 & $ (  1.35  \pm 0.21) \times 10^{-10} $ \\ 
7075 & 1247 & $ (  1.25  \pm  0.20) \times 10^{-10} $ \\ 
7125 & 1163 & $ (  1.18  \pm 0.19) \times 10^{-10} $ \\ 
7175 & 1111 & $ (  1.13  \pm 0.18) \times 10^{-10} $ \\ 
7225 & 1043 & $ (  1.07  \pm 0.17) \times 10^{-10} $ \\ 
7275 & 910 & $ (   9.40  \pm  1.50) \times 10^{-11} $ \\ 
7325 & 897 & $ (  9.33  \pm  1.50) \times 10^{-11} $ \\ 
7375 & 830 & $ (  8.69  \pm  1.40) \times 10^{-11} $ \\ 
7425 & 790 & $ (  8.33  \pm  1.40) \times 10^{-11} $ \\ 
7475 & 758 & $ (  8.05  \pm  1.30) \times 10^{-11} $ \\ 
7525 & 683 & $ (   7.30  \pm  1.20) \times 10^{-11} $ \\ 
7575 & 713 & $ (  7.67  \pm  1.30) \times 10^{-11} $ \\ 
7700 & 2241 & $ (  6.13  \pm    1.00) \times 10^{-11} $ \\ 
7900 & 1791 & $ (  5.03  \pm 0.86) \times 10^{-11} $ \\ 
8100 & 1378 & $ (  3.97  \pm 0.69) \times 10^{-11} $ \\ 
8300 & 1097 & $ (  3.24  \pm 0.58) \times 10^{-11} $ \\ 
8500 & 859 & $ (   2.60  \pm 0.47) \times 10^{-11} $ \\ 
8700 & 670 & $ (  2.08  \pm 0.39) \times 10^{-11} $ \\ 
8900 & 504 & $ (   1.60  \pm 0.31) \times 10^{-11} $ \\ 
9100 & 444 & $ (  1.44  \pm 0.28) \times 10^{-11} $ \\ 
9300 & 328 & $ (  1.09  \pm 0.22) \times 10^{-11} $ \\ 
9500 & 257 & $ (  8.7  \pm  1.8) \times 10^{-12} $ \\ 
9700 & 205 & $ (   7.1  \pm  1.5) \times 10^{-12} $ \\ 
9900 & 183 & $ (  6.5  \pm  1.4) \times 10^{-12} $ \\ 
10250 & 291 & $ (  4.3  \pm 0.9) \times 10^{-12} $ \\ 
10750 & 166 & $ (  2.6  \pm 0.6) \times 10^{-12} $ \\ 
11250 & 100 & $ (  1.6  \pm  0.4) \times 10^{-12} $ \\ 
11750 & 61 & $ (  1.0  \pm 0.3) \times 10^{-12} $ \\ 
12250 & 34 & $ (  6.0  \pm  1.8) \times 10^{-13} $ \\ 
12750 & 31 & $ (  5.7  \pm  1.7) \times 10^{-13} $ \\ 
13250 & 14 & $ (  2.7  \pm 1.0) \times 10^{-13} $ \\ 
13750 & 10 & $ (  2.0  \pm 0.8) \times 10^{-13} $ \\ 
14250 & 11 & $ (  2.3  \pm 0.9) \times 10^{-13} $ \\ 
14750 & 7 & $ (   1.5  \pm 0.7) \times 10^{-13} $ \\ 
15250 & 13 & $ (  2.9  \pm  1.1) \times 10^{-13} $ \\ 
\hline
\end{tabular}
\end{center}
\label{tab:vmuon}
\end{table*}%

\begin{figure}[!ht]
\includegraphics[width = 1.0\columnwidth,keepaspectratio=true,bb=25 25 525 525]{\thedirectory/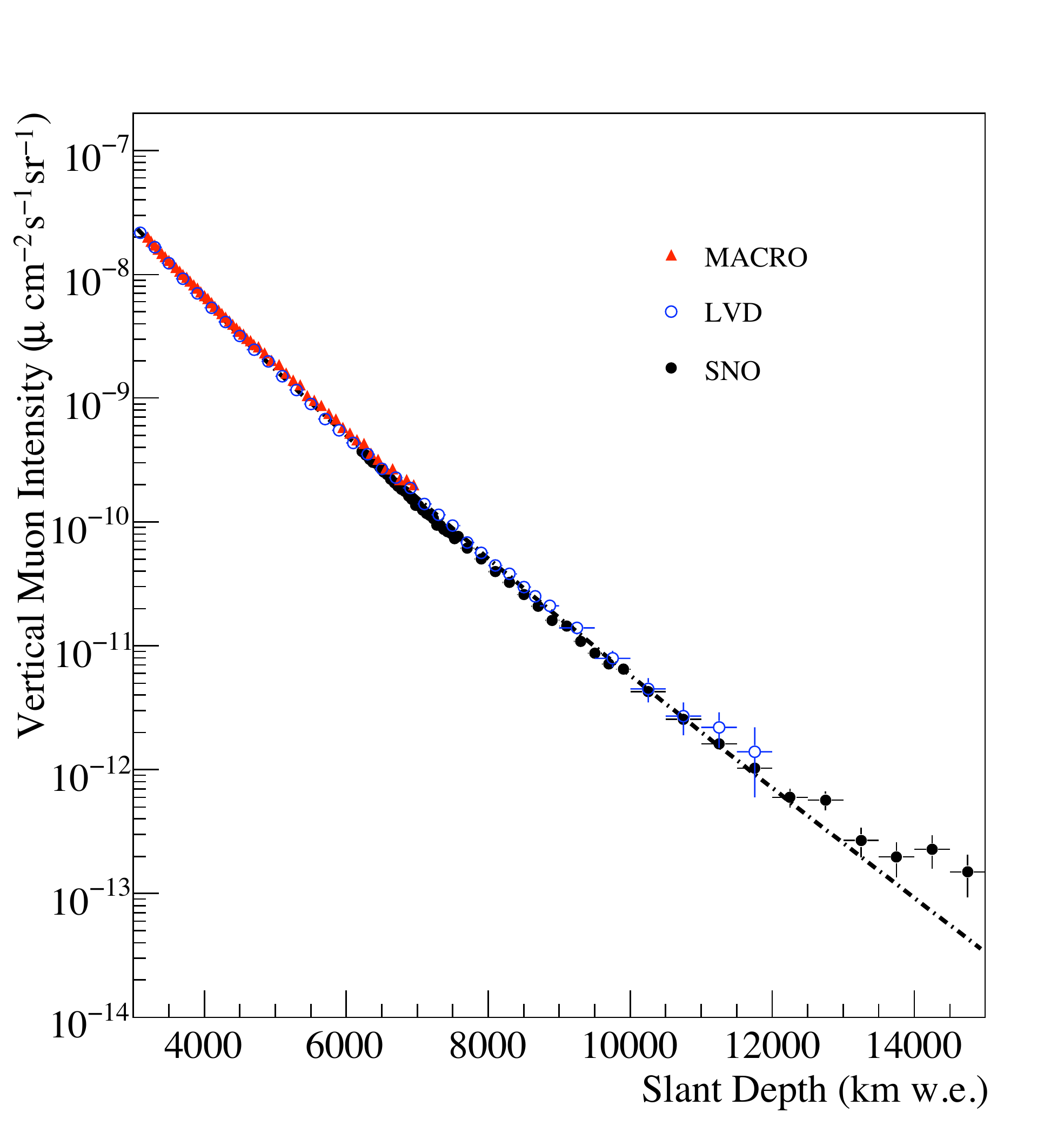}
\caption{The flux of cosmic-ray muons that pass all cuts as a function standard rock depth.  SNO data (filled circles) shown with best global fit intensity distribution (dashed line) and data from LVD~\cite{bib:LVD} (empty circles) and MACRO~\cite{bib:MACRO} (triangles) detectors using Eq.~\ref{eq:attenuation}. Global fit range extends to 13.5 kilometers water equivalent, beyond which atmospheric neutrino-induced muons start to become a significant fraction of the signal.}
\label{fig:muon_flux}
\end{figure}

The attenuation of the vertical muon intensity as a function of depth can be parameterized by:

\begin{equation}
\label{eq:attenuation}
I^v_\mu (x_{\rm std}) = I_0 ~ (\frac{x_0}{x_{\rm std}})^\alpha ~ e^{-x_{\rm std}/x_0} 
\end{equation}

\noindent where $I_0$ is an overall normalization constant, and $x_0$ represents an effective attenuation length for high-energy muons.  The remaining free parameter, $\alpha$, is strongly correlated with the spectral index $\gamma$ in Equation~\ref{eq:surface}.  Results from fits of the vertical muon intensity as a function of depth for various values of these parameters are shown in Table~\ref{tab:vmuon_results}.    We perform fits whereby the parameter $\alpha$ is either fixed to what one would expect from the surface ($\alpha = \gamma -1 = 1.77$) or allowed to float freely.  The cosmic-ray data tends to prefer larger values of $\alpha$ than the expected value of 1.77.  A comparison of SNO's muon flux to that measured in the LVD~\cite{bib:LVD} and MACRO~\cite{bib:MACRO} is shown in Figure~\ref{fig:muon_flux}.  In general, there exists tension between the different data sets.  Fits have been performed both with and without allowing the slant depth uncertainty to float within its uncertainty.  The fits in both cases are nearly identical, with minimal change ($< 1 \sigma$) to the slant depth.  The fits presented in Table~\ref{tab:vmuon_results} are with the slant depth constrained.

To avoid some of the strong correlations between the three parameters listed in Equation~\ref{eq:attenuation}, we also perform the fit using the following parametrization:

\begin{equation}
\label{eq:alternate}
I^v_\mu (x) = e^{(a_0 + a_1 \cdot x + a_2 \cdot x^2)}
\end{equation}

\noindent where $e^{a_0}$ represents the muon flux at the surface, $a_1$ is inversely proportional to the muon attenuation length, and $a_2$ represents the deviation from the simple exponential model.  Results from fitting to Eq.~\ref{eq:alternate} are shown in Table~\ref{tab:alternate}.

The systematic uncertainties of this measurement are summarized in Table~\ref{tab:systematics}.  Certain systematic errors for the cosmic-ray muon flux are in common with those for the neutrino-induced muon flux results, including livetime, impact parameter bias, and angular resolution.  Others are unique to the cosmic-ray muon flux.  These include uncertainties in the rock density, the surface variation, the rock conversion model, muon straggling, instrumental backgrounds, and backgrounds from neutrino-induced events and multiple muons.  This last background is estimated from events measured from the MACRO experiment~\cite{bib:Multiple}.  As the reconstruction for multiple muon events in the detector is not well known, we assign a $\pm 100\%$ uncertainty on this potential background. These systematic uncertainties are included as part of the total error presented in Table~\ref{tab:vmuon_results}.

%
%
\begin{table*}[htdp]
\caption{Results from the SNO fit to the vertical muon intensity for \cosz $> 0.4$ using Equation~\ref{eq:attenuation}.  The fits were performed either using only SNO data with the $\alpha$ parameter allowed to float, with the $\alpha$ parameter fixed to the value predicted from the surface flux of Eq.~\ref{eq:surface} ($\alpha = \gamma - 1 = 1.77$), or combined with LVD~\cite{bib:LVD} and MACRO~\cite{bib:MACRO} cosmic ray data. Symbols in the table are as defined in the text.  The errors reported are a combination of statistical and systematic uncertainties on the flux and slant depth.}
\begin{center}
\begin{tabular}{|c|c|c|c|c|}
\hline
Dataset & $I_0$ & $x_0$& $\alpha$ & $\chi^2/$dof \\
& ($10^{-6}$ cm$^{-2}$ s$^{-1}$ sr$^{-1}$) &  (km w.e.) & & \\
\hline
SNO only & $1.20 \pm 0.69$ & $2.32 \pm 0.27$ & $5.47 \pm 0.38$ & 34.2 / 44 \\
SNO only & $2.31 \pm 0.32$ & $1.09 \pm 0.01$ & $1.77$ & 111.0 / 45 \\
SNO + LVD + MACRO & $2.16 \pm 0.03$ & $1.14 \pm 0.02$ & $1.87 \pm 0.06$ & 230.2/134 \\
\hline
\end{tabular}
\end{center}
\label{tab:vmuon_results}
\end{table*}%

\begin{table*}[htdp]
\caption{Results from the SNO fit to the vertical muon intensity for \cosz $> 0.4$ using Equation~\ref{eq:alternate}.  Fits shown using only SNO data, or combined with LVD~\cite{bib:LVD} and MACRO~\cite{bib:MACRO} cosmic ray data. Symbols in table are as defined in the text.  The errors reported are a combination of statistical and systematic uncertainties on the flux and slant depth.}
\begin{center}
\begin{tabular}{|c|c|c|c|c|}
\hline
Dataset & $e^{a_0}$ & $a_1$& $a_2$ & $\chi^2/$dof \\
& (cm$^{-2}$ s$^{-1}$ sr$^{-1}$) & (m.w.e.)$^{-1}$ & (m.w.e.)$^{-2}$& \\
\hline
SNO only & $(4.55^{+0.90}_{-0.75}) \times 10^{-6}$ & $(-1.75 \pm 0.06) \times 10^{-3}$ & $(3.9 \pm 0.3) \times 10^{-8}$ & 41.6 / 44 \\
SNO + LVD + MACRO & $(1.97 \pm 0.06) \times 10^{-6}$ & $(-1.55 \pm 0.01) \times 10^{-3}$ & $(2.78 \pm 0.08) \times 10^{-8}$ &  230.8 / 134 \\
\hline
\end{tabular}
\end{center}
\label{tab:alternate}
\end{table*}%

\subsection{Atmospheric Neutrino Results}

We assume a model for the atmospheric neutrino flux, and fit for a total flux scaling factor as well as the atmospheric neutrino oscillation parameters.  In these fits we use a two-neutrino mixing model:

\begin{widetext}
\begin{equation}
\label{eq:oscillations}
\Phi(L/E_\nu, \theta, \Delta m^2)_{\mu} = \Phi_0 \cdot [1-\sin^2{2\theta} \cdot \sin^2{(\frac{1.27 \Delta m^2 L}{E_\nu})}]
\end{equation}
\end{widetext}

\noindent where $ \theta$ is the neutrino mixing angle, $\Delta m^2$ is the square mass difference in eV$^2$, $L$ is the distance traveled by the neutrino in km, $E_\nu$ is the neutrino energy in GeV, and $\Phi_0$ is the overall normalization of the neutrino-induced flux.  

Although the signal uncertainty is dominated by statistics, systematic errors do have an impact on both the acceptance and zenith angle distribution of events.  To account for distortions in the zenith angle spectrum, we generalize the $\chi^2$-pull technique (see~\cite{bib:Lisi} and references therein) to the case of a maximum likelihood analysis.  This allows us to account for the smallness of statistics while still incorporating any correlations that may exist between different systematic error contributions. An extended likelihood function is constructed using the following equation:

\begin{equation}
{\cal L}_{\rm total} = 2 (\sum^{N_{\rm bins}}_{i} \ln{(\frac{N^{\rm data}_{i}}{N^{\rm MC}_{i}})} - (N^{\rm MC}_{i} -N^{\rm data}_{i}));
\end{equation}

\noindent where $N^{\rm data(MC)}$ represents the number of data (Monte Carlo) events found in a given zenith bin $i$.  To account for the effect of systematic errors on our likelihood contours, we perform a linear expansion of $N^{\rm MC}$ with respect to a nuisance parameter $\vec{\alpha}$ for each systematic uncertainty such that:

\begin{equation}
N^{\rm MC}_i \simeq N^{\rm MC}_{0,i} + \sum_j^{N_{\rm sys}} (\frac{\partial N^{\rm MC}_{0,i}}{\partial \alpha_j}) \Delta \alpha_j = N^{\rm MC}_{0,i} (1 + \vec{\beta}_i \cdot \Delta \vec{\alpha})
\end{equation}

Note that we have used vector notation to denote a summation over all nuisance parameters.  By expanding the logarithmic term to second order and minimizing the likelihood function with respect to each nuisance parameter, one finds an analytical expression~\cite{bib:Miles}:

\begin{widetext}
\begin{equation}
{\cal L}_{\rm total} = 2 (\sum^{N_{\rm bins}}_{i} \ln{(\frac{N^{\rm data}_{i}}{N^{\rm MC}_{0,i}})} - (N^{\rm MC}_{0,i} -N^{\rm data}_{i})) -\Delta\vec{\alpha}^T_{min} {\bf {\cal S}^{2}}\Delta\vec{\alpha}_{min}
\label{eq:like}
\end{equation}
\end{widetext}

\noindent where $\Delta \vec{\alpha}^T_{min}$ represents the minimized nuisance parameter:

\begin{equation}
\label{eqn:like_simple}
\Delta \vec{\alpha}_{min} = ( \sum_i^{N_{\rm bins}} (N^{data}_{i} - N^{MC}_{0,i}) \vec{\beta}_i) {\bf {\cal S}^{-2}}
\end{equation}

\noindent and the matrix ${\bf {\cal S}^{2}}$ is defined as:

\begin{equation}
{\bf {\cal S}^2} = {\bf \sigma^{-2}} + \sum_i^{N_{bins}} \vec{\beta}_i
\times \vec{\beta}^T_i
\end{equation}

Here, $\sigma^{-2}$ is the diagonal error matrix whose entries represent the size of the systematic error constraints.  As long as the contribution from the systematic errors is small, the above formalism provides a very efficient method for evaluating the effect of systematic errors while also incorporating constraints from the data.  A total of six systematic uncertainties are fit using this method; five of which (axial mass, quasi-elastic cross-section, resonance cross-section, deep inelastic scattering, and energy loss modeling) have explicit zenith angle dependencies, while the last is flat with respect to the zenith distribution. This uncertainty is a combination of all of the remaining systematic errors and is fit as an overall normalization error.  A summary of all the systematic errors is shown in Table~\ref{tab:systematics}.  

\begin{figure}[!ht]
\includegraphics[width = 1.00\columnwidth,keepaspectratio=true,bb=25 25 525 425]{\thedirectory/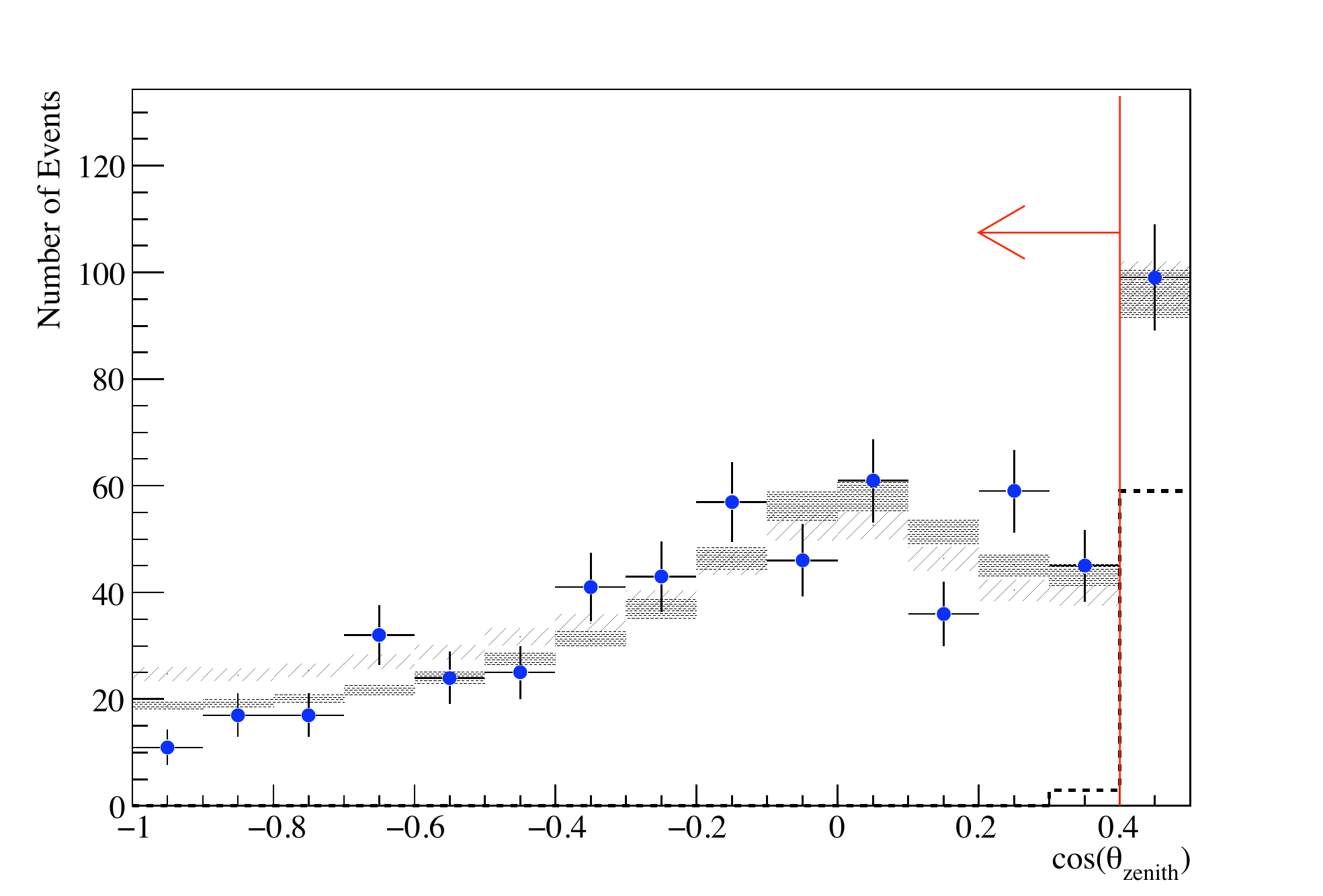}
\caption{The distribution of through-going neutrino-induced muons that pass all cuts as a function of zenith angle.  Data (crosses) are shown with the best-fit MC spectra of $(\Phi_0,\sin^2{2\theta},\Delta m^2)$ = (\normalization, \thetaatm, \msqatm) (solid box) and prediction with no neutrino oscillation and a best fit normalization of $\Phi_0$ = \normalizationnoosc ~(hashed box).  The background due to cosmic-ray muons is shown in the dashed line.  The zenith angle cut is indicated in the figure.}
\label{fig:zenith}
\end{figure}

Figure~\ref{fig:zenith} shows the zenith angle distribution for neutrino-induced muons.  A total of \upwardevents events are recorded with $-1 < $\cosz$ < 0.4$ in the \livetime ~days of livetime in this analysis.  For neutrino-induced events near the horizon ($\cos{\theta_{\rm zenith}}$ between 0 and 0.4), \hupwardevents events are observed. Given the current measurements of the atmospheric oscillation parameters, the neutrino-induced flux is unaffected by oscillations in this latter region and therefore is a direct measurement of the atmospheric neutrino flux, particularly at high energies. The corresponding neutrino-induced through-going muon flux below the horizon (\cosz$ < 0$) and above the horizon ($0 < $\cosz$ < 0.4$) are \totalvflux cm$^{-2}$s$^{-1}$sr$^{-1}$ and \totalhflux cm$^{-2}$s$^{-1}$sr$^{-1}$, respectively.  


\begin{table*}[htdp]
\caption{Summary of systematic errors for the neutrino-induced and cosmic-ray muon flux measurements. A dagger ($\dagger$) indicates that the systematic error only affects the cosmic-ray muon intensity fit to Eq.~\ref{eq:attenuation} and is not included in the total systematic error summation below. The total error in the table is determined from the fit including correlations and does not equal to the quadrature sum of the individual components.}
\begin{center}
\begin{tabular}{|l c|c|c|}
\hline \hline
{\bf Systematic Error} & {\bf Variation} & {\bf $\nu_\mu$-induced muon flux error} & {\bf Cosmic-ray muon flux error} \\
\hline \hline
\multicolumn{4}{|l|}{\bf Detector} \\
\hline
Detector Propagation Model & Various& $\pm 0.3\%$ & $\pm 0.3\%$ \\
Angular Resolution & $\pm 0.6^\circ$ & $\pm 0.1\%$ &  $\pm 0.1\%$ \\
Energy Loss Model & $\pm 5\%$ & $\pm 2.5\%$ & $\pm 0.2\%$\\
Impact Bias/Shift & $\pm 4.0$ cm &  $\pm 1.2\%$ &  $\pm 1.0\%$ \\
Impact Resolution & $\pm 8.5$ cm &  $\pm 0.07\%$ &  $\pm 0.07\%$ \\
Livetime Clock & $\pm 2600$ s & $\pm 0.002\%$ & $\pm 0.002\%$ \\
PMT Charge Model &$\pm 10\%$  & $\pm 1.0\%$ & $\pm 0.05\%$ \\
Fisher Discriminant Cut &$\pm 5\%$ & $\pm 2.1\%$ & $\pm 0.37\%$ \\
\multicolumn{2}{|l|}{\em Total Detector Model} & $\pm 3.7\%$ & $\pm 1.1\%$ \\
\hline
\multicolumn{4}{|l|}{\bf Neutrino Cross-Section Model} \\
\hline
Axial Mass & $\pm 0.15$ GeV & $\pm 1.1\%$ & N/A \\
Quasi-Elastic & $\pm 10\%$ & $\pm 0.8\%$ & N/A \\
Resonance  & $\pm 20\%$ & $\pm 1.9\%$ & N/A \\
Deep Inelastic  & $\pm 3\%$ & $\pm 2.1\%$ & N/A \\
\multicolumn{2}{|l|}{\em Total Cross-Section Model} & $\pm 3.1\%$ & N/A \\
\hline
\multicolumn{4}{|l|}{\bf Muon Propagation Model} \\
\hline
Rock Density($\dagger$) & $\pm 0.05$ g/cm$^3$ & $\pm 0.3\%$ &($\dagger$) \\
Conversion Model($\dagger$) & $\pm 1\%$ &N/A & ($\dagger$) \\
Surface Variation($\dagger$) & $\pm 50$ m & N/A & ($\dagger$) \\
Transport Model &  & $\pm 2\%$ & N/A  \\
Time/Seasonal Variation & & $\pm 1\%$ & $\pm 2.2\%$ \\
\multicolumn{2}{|l|}{\em Total Propagation Model} & $\pm 2.2\%$ & $\pm 2.2\%$ \\
\hline
\multicolumn{4}{|l|}{\bf Backgrounds} \\
\hline
Instrumental & $0.3 \pm 0.2$ events yr$^{-1}$ & $\pm 0.2\% $ & $<0.1\%$ \\
Cosmic ray $\mu$ & $0.6 \pm 1.1$ events yr$^{-1}$& $\pm 0.8\%$ & N/A \\
$\nu_\mu$-Induced & $45.8 \pm 2.3$ events yr$^{-1}$ & N/A & $\pm 0.2 \%$ \\
Multiple Muons & $\pm 100\%$ & $\ll 1\%$ & $\pm 1\%$ \\
\multicolumn{2}{|l|}{\em Total Background Error} & $\pm 0.8\%$ & $\pm 1\%$ \\
\hline
\multicolumn{2}{|l|}{\bf Total Systematic Error} & $\pm 4.8\%$ & $\pm 2.7\%$ \\
\hline
\multicolumn{2}{|l|}{\bf Statistical Error} &$+ 8.5\%$ & $\pm 0.4\%$\\
\hline \hline
\end{tabular}
\end{center}
\label{tab:systematics}
\end{table*}%

From the measured zenith angle distribution, we can extract the flux normalization $\Phi_0$ and the neutrino mixing parameters $\theta$ and $\Delta m^2$ in Equation~\ref{eq:oscillations}.  A maximum likelihood fit is performed to find the best fit points, as outlined above.  If all parameters are allowed to float, one finds a flux normalization value of $\Phi_0=$\normalization~and best fit neutrino oscillation parameters of $\Delta m^2$ of \msqatm and maximal mixing. These results are with respect to the Bartol three-dimensional atmospheric flux model and the cross-section model implemented in NUANCE described in Section~\ref{sec:flux}~\cite{bib:Barr04}.  The zenith angle spectrum is consistent with previously measured neutrino oscillation parameters.  One can also look at SNO's sensitivity on the atmospheric flux $\Phi_0$ by including existing constraints on the atmospheric neutrino oscillation parameters from the Super-Kamiokande~\cite{bib:Ashie05} $(\Delta m^2, \sin^2{2\theta}_{\rm SK}) = (2.1^{+0.6}_{-0.4} \times 10^{-3}{\rm~eV}^2, 1.000 \pm 0.032)$ and MINOS~\cite{bib:Michael06,bib:MINOS08} $(\Delta m^2_{\rm MINOS} = (2.43\pm0.13) \times 10^{-3}{\rm~eV}^2)$ neutrino experiments.  The likelihood function in Eq.~\ref{eq:like} is altered to the following: 

\begin{widetext}
\begin{equation}
{\cal L}_{\rm constrained} = {\cal L}_{\rm total} + (\frac{\Delta m^2 - \Delta m^2_{\rm SK}}{\sigma_{\rm \Delta m^2,SK}})^2 + (\frac{\Delta m^2 - \Delta m^2_{\rm MINOS}}{\sigma_{\rm \Delta m^2,MINOS}})^2 + (\frac{\sin^2{2\theta}-\sin^2{2\theta}_{\rm SK}}{\sigma_{\rm \theta,SK}})^2 
\end{equation}
\end{widetext}

The constraint reduces the uncertainty on the overall atmospheric neutrino flux normalization to \normalizationconst.  The 68\%, 95\% and 99.73\% confidence level regions for the parameters as determined by the fits are shown in Figure~11. The scenario of no neutrino oscillations by using SNO-only data is excluded at the \CL confidence level.

\begin{figure*}[htdp]
\begin{tabular}{c c}
\includegraphics[width = 1.0\columnwidth,keepaspectratio=true,bb=0 0 525 375]{\thedirectory/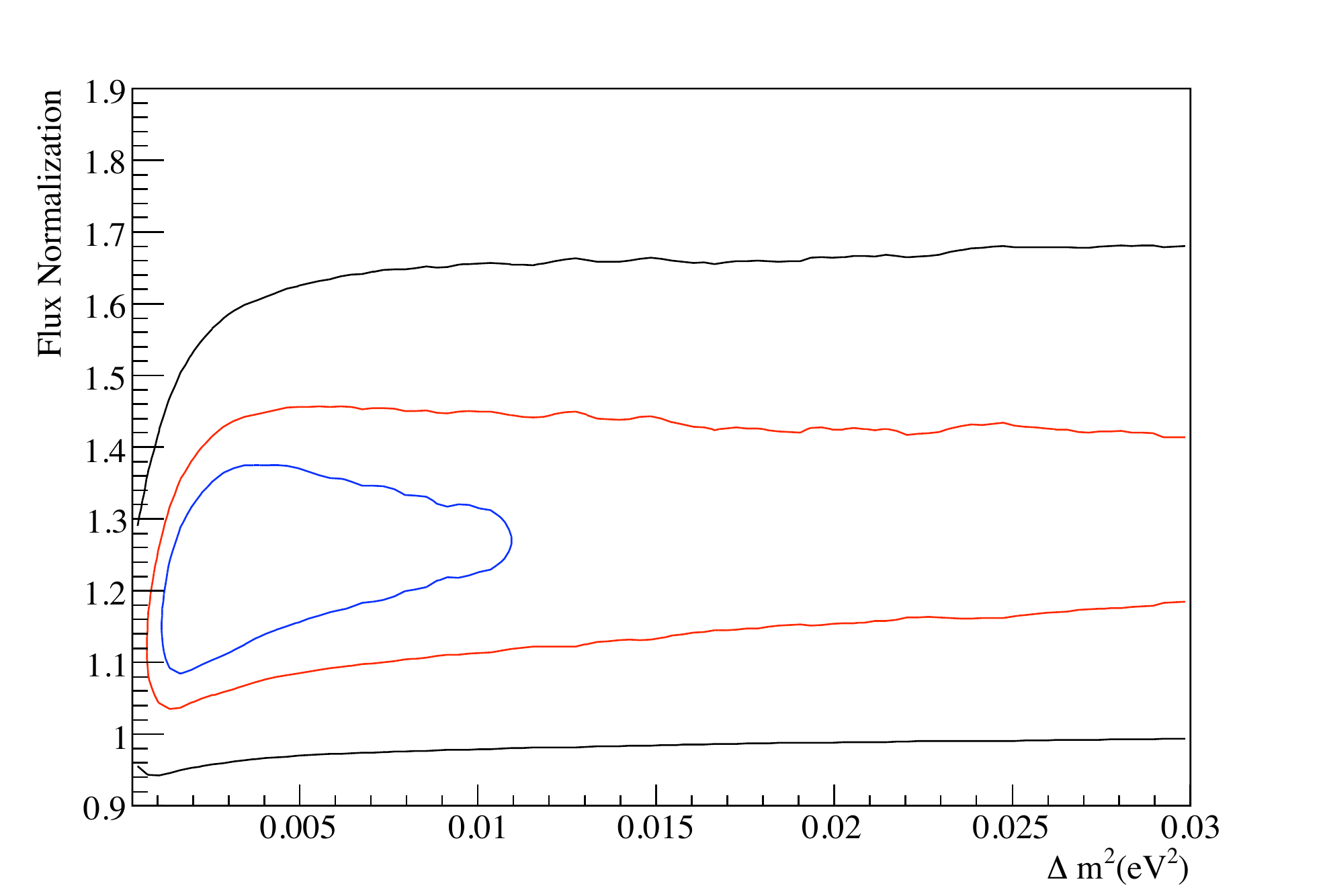} &
\includegraphics[width = 1.0\columnwidth,keepaspectratio=true,bb=0 0 525 375]{\thedirectory/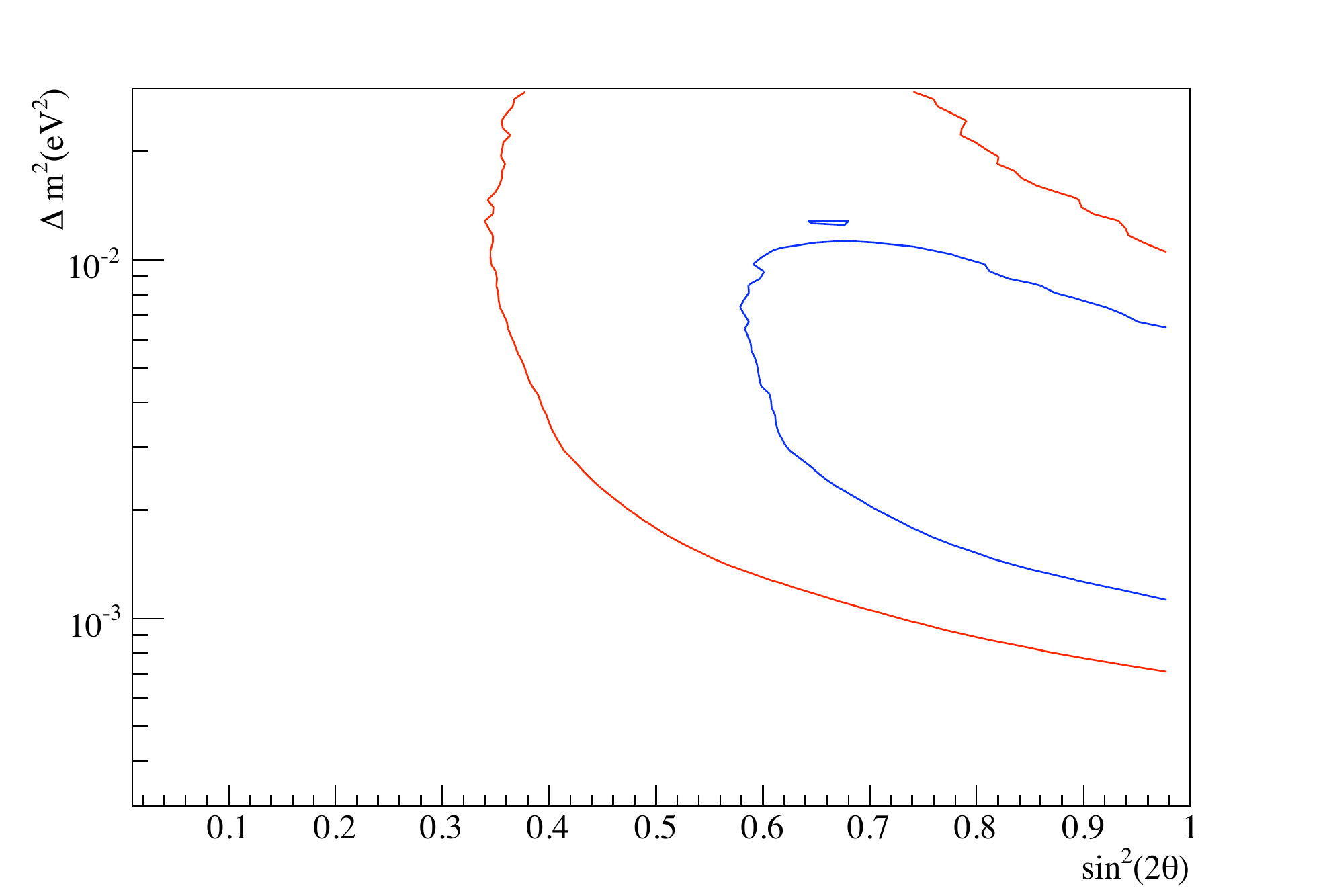} \\
\includegraphics[width = 1.0\columnwidth,keepaspectratio=true,bb=0 0 525 375]{\thedirectory/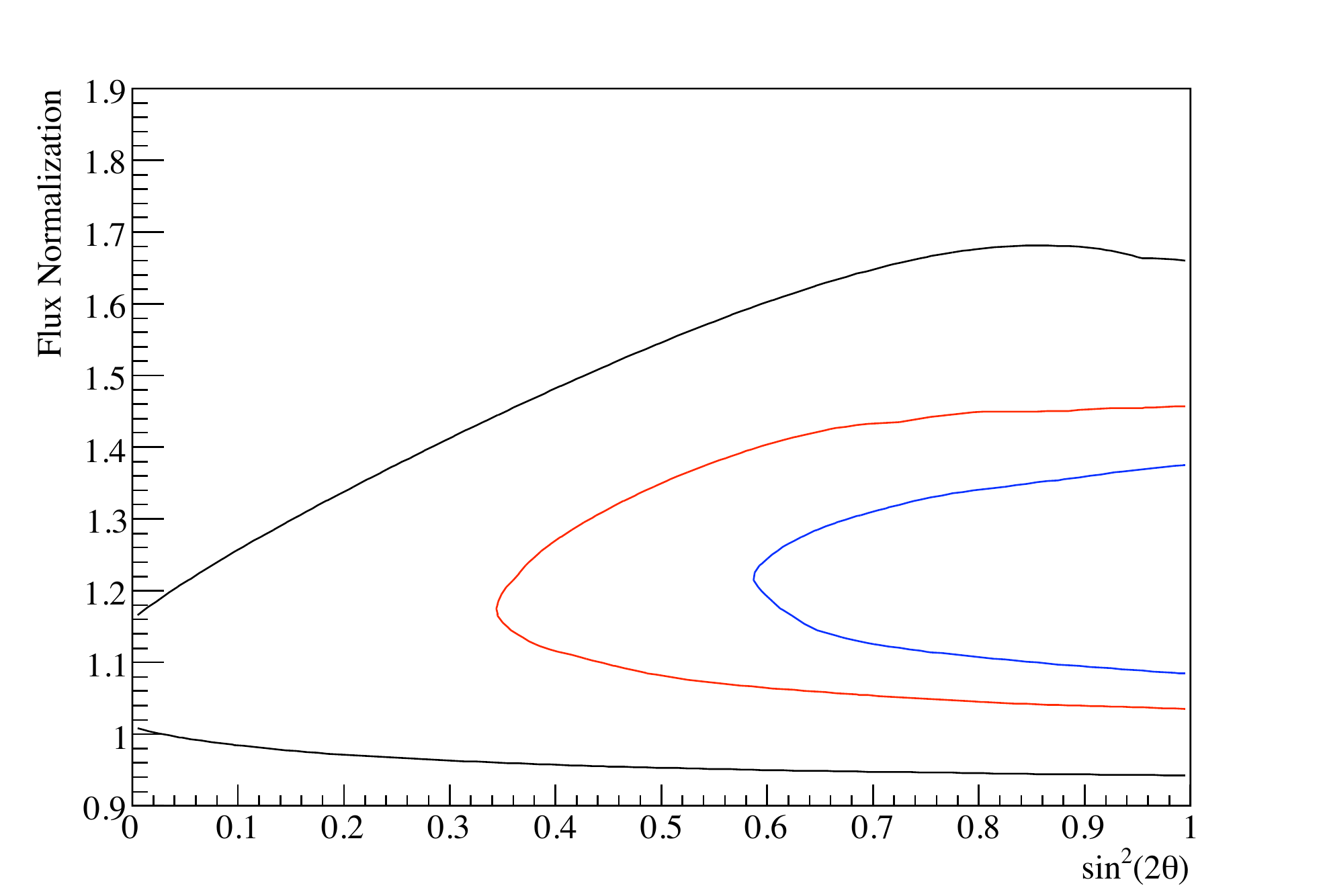} &
\includegraphics[width = 1.0\columnwidth,keepaspectratio=true,bb=0 0 525 375]{\thedirectory/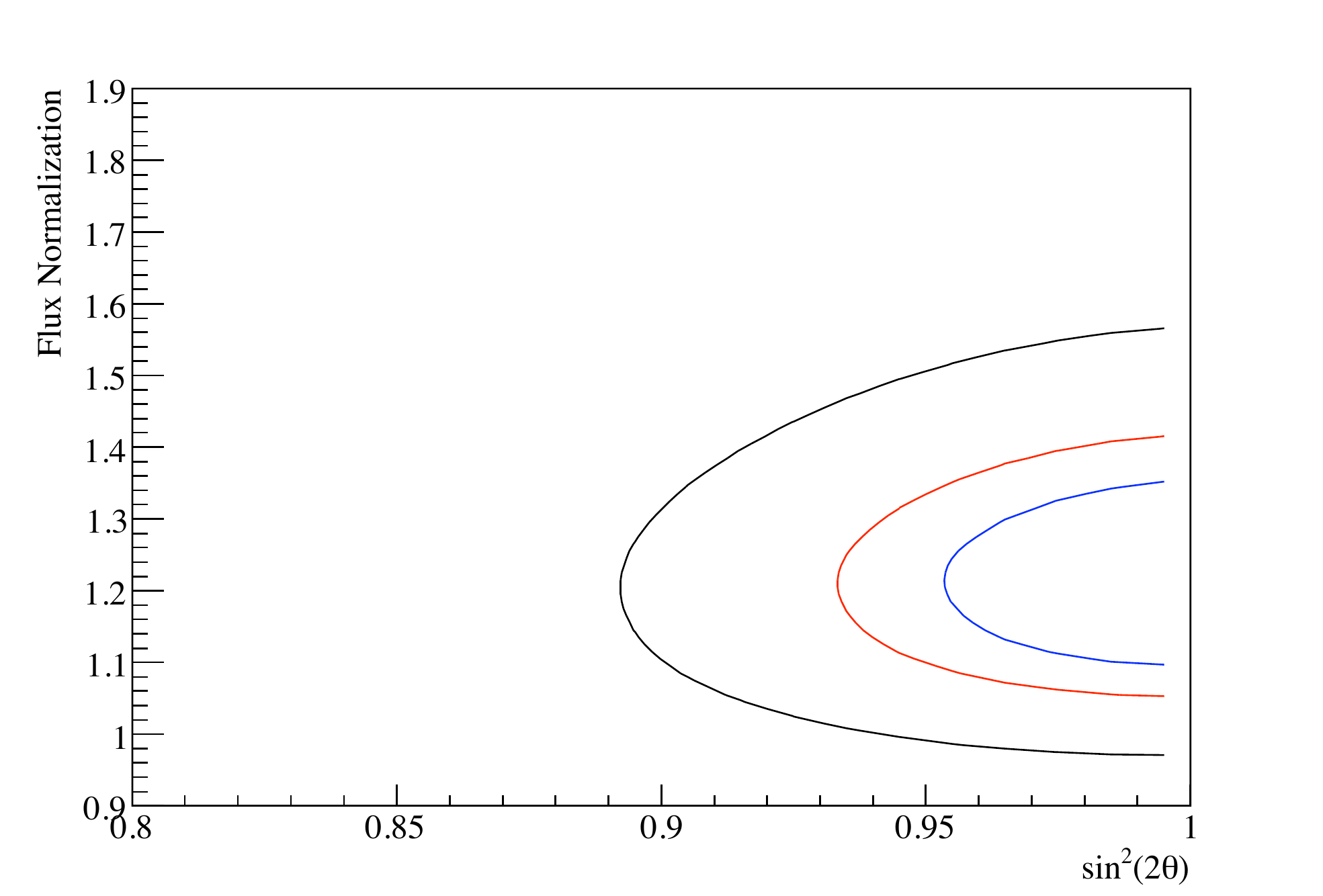} \\
\label{fig:cl}
\end{tabular}
\caption{The 68\% (blue), 95\% (red), and 99.73\% (black) confidence level contours for the $\nu_\mu$ atmospheric neutrino oscillation parameters based on the muon zenith angle distribution for \cosz $<0.4$.  The plots show the SNO-only contours for flux normalization versus mass splitting (top left), SNO-only mass splitting versus mixing angle (top right), SNO-only contours for flux normalization versus mixing angle (bottom left) and the flux normalization versus mixing angle including constraints from the Super-K and MINOS neutrino oscillation experiments (bottom right)~\cite{bib:Michael06, bib:Ashie05}.}
\end{figure*}

\section{Summary}
\label{sec:summary}

The Sudbury Neutrino Observatory experiment has measured the through-going muon flux at a depth of 5890 meters water equivalent. We find the total muon cosmic-ray flux at this depth to be \flux. We measure the through-going muon flux induced by atmospheric neutrinos. The zenith angle distribution of events rules out the case of no neutrino oscillations at the $3\sigma$ level. We measure the overall flux normalization to be \normalizationconst, which is larger than predicted from the Bartol atmospheric neutrino flux model but consistent within the uncertainties expected from neutrino flux models.  This is the first measurement of the neutrino-induced flux above the horizon in the angular regime where neutrino oscillations are not an important effect.  The data reported in this paper can be used to help constrain such models in the future.

\section{Acknowledgements}
\label{sec:ack}

This research was supported by: Canada: Natural Sciences and Engineering Research Council, Industry Canada, National Research Council, Northern Ontario Heritage Fund, Atomic Energy of Canada, Ltd., Ontario Power Generation, High Performance Computing Virtual Laboratory, Canada Foundation for Innovation; US: Dept. of Energy, National Energy Research Scientific Computing Center; UK: Science and Technology Facilities Council; Portugal: Funda\c{c}\~{a}o para a Ci\^{e}ncia e a Tecnologia. We thank the SNO technical staff for their strong contributions. We thank Vale Inco for hosting this project.

\end{document}